\documentclass[manuscript]{emulateapj-rtx4}
\usepackage{natbib,amsmath}
\begin{document}
\title{CONSTRAINING TIDAL DISSIPATION IN STARS FROM THE DESTRUCTION RATES OF
EXOPLANETS}
\author{Kaloyan Penev}
\affil{Department of Astrophysical Sciences, 4 Ivy Lane, Peyton Hall,
Princeton University, Princeton, NJ 08544, USA}
\author{Brian Jackson}
\affil{Carnegie DTM, 5241 Broad Branch Road, NW, Washington, DC 20015-1305,
USA}
\author{Federico Spada}
\affil{Department of Astronomy, Yale University, P.O. Box 208101, New Haven,
CT 06520-8101, USA}
\author{Nicole Thom}
\affil{Nasa Goddard Space Flight Center, Greenbelt, MD, USA}

\begin{abstract}
	We use the distribution of extrasolar planets in circular orbits
	around stars with surface convective zones detected
	by ground based transit searches to constrain how efficiently tides
	raised by the planet are dissipated on the parent star. We
	parameterize this efficiency as a tidal quality factor ($Q_*$). We
	conclude that the population of currently known planets is
	inconsistent with $Q_*<10^7$ at the 99\% level. Previous studies show
	that values of $Q_*$ between $10^5$ and $10^7$ are required in order
	to explain the orbital circularization of main sequence low mass
	binary stars in clusters, suggesting that different dissipation
	mechanisms might be acting in the two cases, most likely due to the
	very different tidal forcing frequencies relative to the stellar
	rotation frequency occurring for star--star versus planet--star
	systems.
\end{abstract}

\keywords{convectiov --- planet--star interactions --- stars: interiors ---
stars: rotation --- stars: winds, outflows --- turbulence}
\section{Introduction}

Exoplanets with orbital distances $\le$ 0.1 AU from their host stars,
called close-in exoplanets, have presented an especially puzzling challenge
to theories of planet formation. The protoplanetary disk is too warm (~2000
K) so close to a star to allow the condensation and accumulation of icy and
rocky material required to form planets \citep{Lin_et_al_96, Miller_et_al_09,
Ibgui_Burrows_09}. \\


%
For close-in exoplanetary systems, their mutual tidal gravities
induce significant tidal bulges in the planets and stars. Dissipation of the
accompanying tidal energy drives obliquities to zero and rotation rates to
near synchronous, processes that probably take millions of years for
planets but billions of years for stars. While its orbit is eccentric,
dissipation of tidal energy within a planet can reduce the orbital semi-major
axis and eccentricity, as well as warming the planet's interior with
significant consequences for the planet's thermal evolution
\citep{Jackson_et_al_08b, Ibgui_Burrows_09, Liu_et_al_08, Miller_et_al_09}.
The majority of planet-hosting stars rotate more slowly than their close-in
planets revolve, and so tides raised on these stars also reduce
eccentricities and semi-major axes. \\



Although the effects of tides raised on close-in exoplanets become
negligible as eccentricities shrink, tides on host stars continue to reduce
semi-major axes long after eccentricities are negligible, as long as the
stellar rotation rate is smaller than the orbital mean motion. For the
systems for which the rotation of the star is observationally constrained,
the stellar spin period is known to be longer than the orbital period.
Typically, planet--hosting stars are older than 1 Gyr, and so stars without
reported rotation rates likely rotate slower than their planets revolve.
Moreover, observational biases favor detection of planets around slow
rotators. As a result, the tides on the host star tend to
dominate the long-term tidal evolution of close-in planets. Eventually, the
planets may cross their Roche limits (0.007 AU for a Jupiter-like planet
around a Sun-like star), where they are tidally disrupted.\\


On the other hand, tidal spin-up may synchronize the stellar rotation
to the orbital period, in which case the planet will eventually reach a
stable orbit. The total angular momentum of the system determines which
scenario occurs \citep{Counselman_73, Greenberg_74}.
\citet{Levrard_et_al_09} show that all systems with transiting
planets found to date by ground transit search surveys except HAT-P 2 b have
insufficient angular momentum to prevent this destruction. Loss of angular
momentum through shedding of stellar wind dooms even HAT-P 2 b. Thus, given
enough time, the loss of close-in exoplanets through orbital decay is
inevitable.\\

Tidal evolution of an orbit increases rapidly for decreasing
semi-major axis, and so the distribution of semi-major axes for observed
planets is sensitive to the rate of tidal dissipation. Since the probability
for a planet to transit its host star increases for planets nearer their
stars, transiting planets are especially susceptible to tidal effects. The
rate of orbital decay and frequency of tidal destruction also depends
sensitively on the rate of tidal dissipation within the host star. This rate
is related to the efficiency parameter $Q_*$ \citep{Goldreich_63}: larger
$Q_*$ corresponds to less tidal dissipation and slower orbital evolution. The
origins of tidal dissipation within gaseous planets and stars have been
studied extensively, but remain poorly understood, with estimates based on
theoretical and observational studies ranging from $10^5$ to $10^9$
\citep[c.f.][]{Zahn_66, Zahn_70, Zahn_75, Zahn_77, Zahn_89,
Goldreich_Nicholson_77, Scharlemann_81, Scharlemann_82, Goodman_Oh_97,
Savonije_Papaloizou_97a, Savonije_Papaloizou_97b, Savonije_Papaloizou_97c,
Terquem_et_al_98, Ogilvie_Lin_04, Ogilvie_Lin_07, Wu_05a, Wu_05b,
Papaloizou_Ivanov_05, Meibom_Mathieu_05, Ivanov_Papaloizou_07,
Penev_Sasselov_Robinson_Demarque_07, Penev_Barranco_Sasselov_09,
Penev_Sasselov_Robinson_Demarque_09, Penev_Barranco_Sasselov_11, Ogilvie_09,
Penev_Sasselov_11}.\\

Consequently, the time until a planet crosses its Roche limit and is
removed, which we will call a planet's time left (TL), depends both on $Q_*$
and on its current semi-major axis, among other parameters. For a population
of tidally evolving planets, we expect to find few planets with TL much
less than the whole lifetime or current system age. Otherwise, we would
conclude that we have caught a large fraction of the planets in the last
extremely short moments of their lives, just before they are disrupted by
their star. By tuning $Q_*$ until we generate a statistically likely
distribution of TL, we can constrain $Q_*$ and the frequency of tidal
disruption of exoplanets. \\

As discussed in Section 2, observational biases have important and complex
influences on the distribution of calculated TL-values for transiting planets
and must be considered in order to produce statistically reliable constraints
on $Q_*$. Several previous studies \citep[e.g.,][]{Carone_Patzold_07} have
attempted to place constraints on $Q_*$ using considerations similar to ours
but only applied to individual planets. Results from some of those studies
are consistent with $Q_* > 10^7$, but inferences based on the orbital
evolution of a single planet may not be statistically meaningful.

The outline of this paper is as follows: in Section \ref{sec: orbital
evolution} we describe our methods and assumptions for calculating the
orbital evolution of planetary systems; in Section \ref{sec: list of systems}
we show the sample of transiting planets and the corresponding parameters
which were used in this work; in Section \ref{sec: biases} we
discuss the various observational and astrophysical biases that affect the
sample of systems with transiting planets found by ground based transit
searches and our procedures and assumptions for how to deal with them in the
analysis; in Section \ref{sec: testing Q*} we outline the procedure we use to
derive constraints on the $Q_*$ value; in Section \ref{sec: results} we show our
limits to $Q_*$; in Section \ref{sec: discussion} we show the apparent
discrepancy between our results and estimates of $Q_*$ derived from binary
stars in open clusters. 

\section{Orbital Evolution}
\label{sec: orbital evolution}
Tidal decay of close-in planets involves the exchange of angular momentum
between a planet's orbit and its host star's rotation. For the stars we
consider here, several processes influence the stellar rotation, in addition
to tidal processes, and accurate modeling of the orbital decay requires
consideration of these effects. For this purpose, we solve the following
system of ordinary differential equations:
\begin{eqnarray}
	\frac{da}{dt}=\mathrm{sign}(\omega_\mathrm{conv}&-&\omega_\mathrm{orb})
	\frac{9}{2}\sqrt{\frac{G}{aM_*}}\left( \frac{R_*}{a}
	\right)^5\frac{m_p}{Q_*}\quad \label{eq: adot}\\
	\left(\frac{dL_\mathrm{conv}}{dt}\right)_\mathrm{tide}&=&-\frac{1}{2}m_p
	M_*\sqrt{\frac{G}{a(M_*+m_p)}}\frac{da}{dt} \label{eq: tidalLdot}\\
	\left(\frac{dL_\mathrm{conv}}{dt}\right)_\mathrm{wind}&=&-K\omega_\mathrm{conv} 
	\min(\omega_\mathrm{conv}, \omega_\mathrm{sat})^2 \\
	&&\left(
	\frac{R_*}{R_\odot}\right)^{1/2} \left( \frac{M_*}{M_\odot}
	\right)^{-1/2}\\
	\label{eq: wind}
	\frac{dL_\mathrm{conv}}{dt}&=&\frac{\Delta L}{\tau_c} - \frac{2}{3}
	R_\mathrm{rad}^2
	\omega_\mathrm{conv} \frac{dM_\mathrm{rad}}{dt} + \nonumber\\
	&&{}+\left(\frac{dL}{dt}\right)_\mathrm{wind} +
	\left(\frac{dL}{dt}\right)_\mathrm{tide} \label{eq: convLdot}\\
	\frac{dL_\mathrm{rad}}{dt}&=&-\frac{\Delta L}{\tau_c} + \frac{2}{3}
	R_\mathrm{rad}^2
	\omega_\mathrm{conv} \frac{dM_\mathrm{rad}}{dt}\label{eq: radLdot}\\
	\Delta
	L&=&\frac{I_\mathrm{conv}L_\mathrm{rad}-I_\mathrm{rad}L_\mathrm{conv}}{I_\mathrm{conv}+I_\mathrm{rad}}
	\label{eq: deltaL} \label{eq: orbital evol last}
\end{eqnarray}
where $M_*$ is the mass of the star; $R_*$ is the radius of the star; $m_p$
is the mass of the planet; $Q_*$ is the tidal quality factor of the star; 
$\mathrm{sign}(\omega_\mathrm{conv}-\omega_\mathrm{orb})$ takes the value 1
when the stellar convective zone is spinning faster than the planet and -1
when the reverse is true; $K=0.35\ M_{\odot}R_{\odot}^2
\mathrm{day}^2\mathrm{Gyr}^{-1}$ is the proportionality constant,
parametrizing the strength of the magnetic wind of the star;
$\omega_\mathrm{sat}=1.84\ \mathrm{day}^{-1}$ is the wind saturation
frequency; $I_\mathrm{conv}$ is the moment of inertia of the stellar
convective zone; $L_\mathrm{conv}$ is the angular momentum of the stellar
convective zone; $I_\mathrm{rad}$ is the moment of inertia of the stellar
radiative core; $L_\mathrm{rad}$ is the angular momentum of the stellar
radiative core; $\tau_c=5\ \mathrm{Myr}$ is the stellar core-envelope
coupling timescale; $M_\mathrm{rad}$ is the mass of the stellar radiative
core; $R_\mathrm{rad}$ is the radius of the radiative-convective boundary in
the star; $\omega_\mathrm{conv}\equiv L_\mathrm{conv}/I_\mathrm{conv}$ is the
angular frequency of the stellar convective zone.

We wish to follow a planet--star system as its semi--major axis shrinks under the
influence of tidal friction. There are two sources of friction: the tides on the
star and those on the planet. However, the latter is only important as long as
either the orbit is eccentric and/or the planet is rotating asynchronously.
Since the angular momentum stored in the rotation of the planet is quite small
compared to the orbital or stellar spin angular momenta, it is safe to assume
that the planet spin is synchronized quickly compared to any orbital evolution.
Further, we will restrict our sample to only systems with nearly circular
orbits. In this case, the evolution of the semi-major axis ($a$) is given by
Equation (\ref{eq: adot}) above \citep{Goldreich_63, Kaula_68,
Jackson_et_al_08a}. This expression makes the approximation that the planet's
mass can be neglected compared to the star's mass, a perfectly reasonable
assumption for all the systems we consider, given the uncertainty in the
value of $Q_*$.

The angular momentum that is taken away from the orbit by the tidal
friction is deposited in the star, acting to spin it up, and while for most
of the known transiting planets the orbit does not have enough angular
momentum to spin up the star to synchronous rotation, for at least one planet
this is not true. Further, since the tidal friction couples the planet with
the surface convective zone of the star, it is possible that if the
core-envelope coupling is not sufficient, synchronous rotation can be imposed
on the envelope only. For this reason we also follow the evolution of the
angular momentum of the star ($L_*$) -- Equation (\ref{eq: tidalLdot}). Here,
we do not make the approximation $M_*\gg m_p$, like we did for the orbital
evolution above, in order to make the final set of equations conserve angular
momentum exactly.\\

In addition, since we are interested in timescales of order Gyrs, we cannot
neglect the spin-down of the host star due to its own magnetic wind. The
effects of stellar wind shedding on rotation have been extensively studied, but
remain poorly understood. Equation (\ref{eq: wind}) represents the current
best description of these effects and is motivated by a combination of theory
and observation \citep{Stauffer_Hartmann_87, Kawaler_88, Barnes_Sofia_96}.
Note that we do not introduce any free parameters to describe the stellar
wind. The value of $K$ in Equation (\ref{eq: wind}) is determined by the present
rotation rate and age of the Sun and the value of the wind saturation
frequency $\omega_\mathrm{sat}$ comes from fitting the observations of stellar
rotation rates in open clusters of different ages. With those values, the
expression above matches the rotational evolution of single stars in open
clusters at young ages, as well as the rotation rate of the Sun and other less
well constrained but older stellar populations. Hence, theoretical
interpretations aside, it can be viewed as a parameterization of the rotation
rate observations over the range of ages we encounter during the orbital
evolution of the exoplanet systems in our sample.\\

Open cluster rotation rates impose one more complication on our model:
core-envelope de-coupling. In order to explain how stars with a wide spread of
rotation rates at young ages end up with similar rotation rates later, it is
necessary to allow for quickly rotating cores in the slow surface rotation
rate stars, and a re-distribution of the excess angular momentum to the
convective zone at a later time \citep{Irwin_et_al_07, Irwin_Bouvier_09,
Denissenkov_10}. The expressions for the separate core and envelope evolution
were derived by \citet{Allain_98}, and for planet hosts take the form of
Equations (\ref{eq: convLdot}) and (\ref{eq: radLdot}). We test the
sensitivity of our results on the assumed core-envelope coupling by repeating
our analysis under the assumption that the core and envelope are perfectly
coupled ($\tau_c=0$).\\

Finally, Equation (\ref{eq: deltaL}) for the stellar core--envelope
differential rotation was proposed by \citet{MacGregor_91}.\\

Note that in Equation (\ref{eq: adot}) we treat $Q_*$ as having a fixed
magnitude. We feel that more complex assumptions are not justified, since
the dependence of $Q_*$  on frequency is poorly known, and there is
substantial disagreement between observational and theoretical estimates as
discussed in the introduction.\\

In this work we start a planetary system's evolution from an age of 5 Myr
after the (model) birth of the star, with an initial orbital separation that
evolves to the observed semi-major axis at the present system age. The
initial age is assumed to be early enough so that no significant spin up of
the star due to the tides raised by the planet has occurred, but any
non-tidal evolution of the planetary orbit has stopped. This is reasonable if
one assumes that hot Jupiters arrive at their extremely close-in orbits
through disk migration, since by that time the protoplanetary disk has
dissipated \citep{Haisch_Lada_Lada_01, Bouwman_et_al_06}, but for other
migration mechanisms this might not be the case.\\

The advantage of starting the evolution at 5 Myr is that with these
assumptions the initial stellar rotation distribution is relatively well
constrained from observations of young open clusters \citep[and references
therein]{Irwin_Bouvier_09}. However, this age is significantly before the
star has arrived at the main sequence, so all stellar parameters, in
particular the radius, the masses of the radiative core and convective
envelope, and the corresponding moments of inertia all significantly evolve
with time. This forces us to allow for stellar evolution along with
the orbital and spin evolution of the planet--star systems.  In order to
follow the evolution of the star, we use evolution tracks calculated using
the YREC stellar evolution code \citep{Demarque_et_al_08} with masses
0.4$M_\odot$, 0.5$M_\odot$, 0.6$M_\odot$, 0.7$M_\odot$, 0.8$M_\odot$,
0.9$M_\odot$, 1.0$M_\odot$, 1.05$M_\odot$, 1.1$M_\odot$ and 1.25$M_\odot$.
The evolution of individual stars is determined from cubic spline
interpolation within this grid of models. Stars below 0.4$M_\odot$ and above
1.25$M_\odot$ are excluded from our analysis (see below).\\

In our evolution we assume that the tides couple the orbit of the planet only
to the convective zone. While, in general, we should split the tidal torque in
two parts, one spinning up the convective zone and the other spinning up the
core, the coupling to the core is likely to be negligible compared
to the convective zone for two reasons: (1) the amplitude of the tidal
deformation scales strongly with radius, so it will be much smaller for the
core than for the convective zone; (2) currently there is no known mechanism
for dissipating the tidal energy in the core, so averaged over an orbit there
should be no net angular momentum transfer from the tides to the core.\\

An additional complication for our models is that the prescription described
above is only valid for low stellar masses. For masses larger than
approximately 1.2$M_\odot$ the surface convective zone becomes negligible in
mass, so we cannot treat it as the only sink for angular momentum. Further,
\citet{Wolff_Simon_97} indicate that the angular momentum loss described
above is only valid for low--mass stars ($M_*\lesssim1.3M_\odot$). For more
massive stars, presumably the lack of surface convection suppresses the
stellar wind, and the loss of angular momentum is much weaker. For reasons of
numerical stability, for stars above $1.1M_\odot$ we ignore the core--envelope
decoupling, treating the star as a solid body and we completely exclude from
our analysis stars above $1.25M_\odot$, since those stars do not posses a
significant surface convection zone and could be subject to a different mode
of tidal dissipation, in addition to weaker stellar wind. Consequently,
close--in planets around very massive stars may have much larger TL values
than around less massive stars, with possible implications for planet surveys
of massive stars.\\

One particular effect of including the stellar wind for low--mass stars is
that even after the star has synchronized its spin with the orbit, the
semi--major axis continues to decay. 
In fact, a positive feedback loop is created at
this point: tides keep the stellar spin synchronized with the planet,
so as the stellar wind removes angular momentum, the orbit continues
to shrink. This spins up the star, enhancing the stellar wind, which
draws more angular momentum from the system and increases the rate of
orbital decay. The effect is that after a short amount of time the tidal
spin--up fails to keep up with the angular momentum lost to the wind and the
spin--orbit lock is lost.\\

It should be noted that  the above tidal evolution equations 
are only valid under the assumption of good alignment between the orbital and
stellar angular momenta. This has been checked only for a subset of the
currently known transiting extrasolar planets, through the measurement of the
Rossiter--McLaughlin effect, and a non-trivial fraction of misaligned planets
is found \citep[etc.]{Johnson_et_al_11, Winn_et_al_11,
Narita_et_al_10a, Narita_et_al_10b, Winn_et_al_10a, Winn_et_al_10b,
Bayliss_et_al_10, Tripathi_10, Johnson_et_al_09, Winn_et_al_09,
Johnson_et_al_08, Winn_et_al_08, Narita_et_al_08, Narita_et_al_07,
Winn_et_al_06, Queloz_et_al_00, Bundy_Marcy_00, Snellen_04}. However, even
for significant misalignment, the tidal torques will only change by a
factor of order unity, much smaller than the uncertainties on $Q_*$, which
range over several orders of magnitude. In addition, the current
measurements of the stellar spin--orbit alignment suggest that the shortest
period planets, which are the only ones sensitive to the tidal dissipation,
are well aligned with their parent star's rotation.\\

\begin{figure*}
	\begin{center}
		\includegraphics[width=1.0\textwidth]{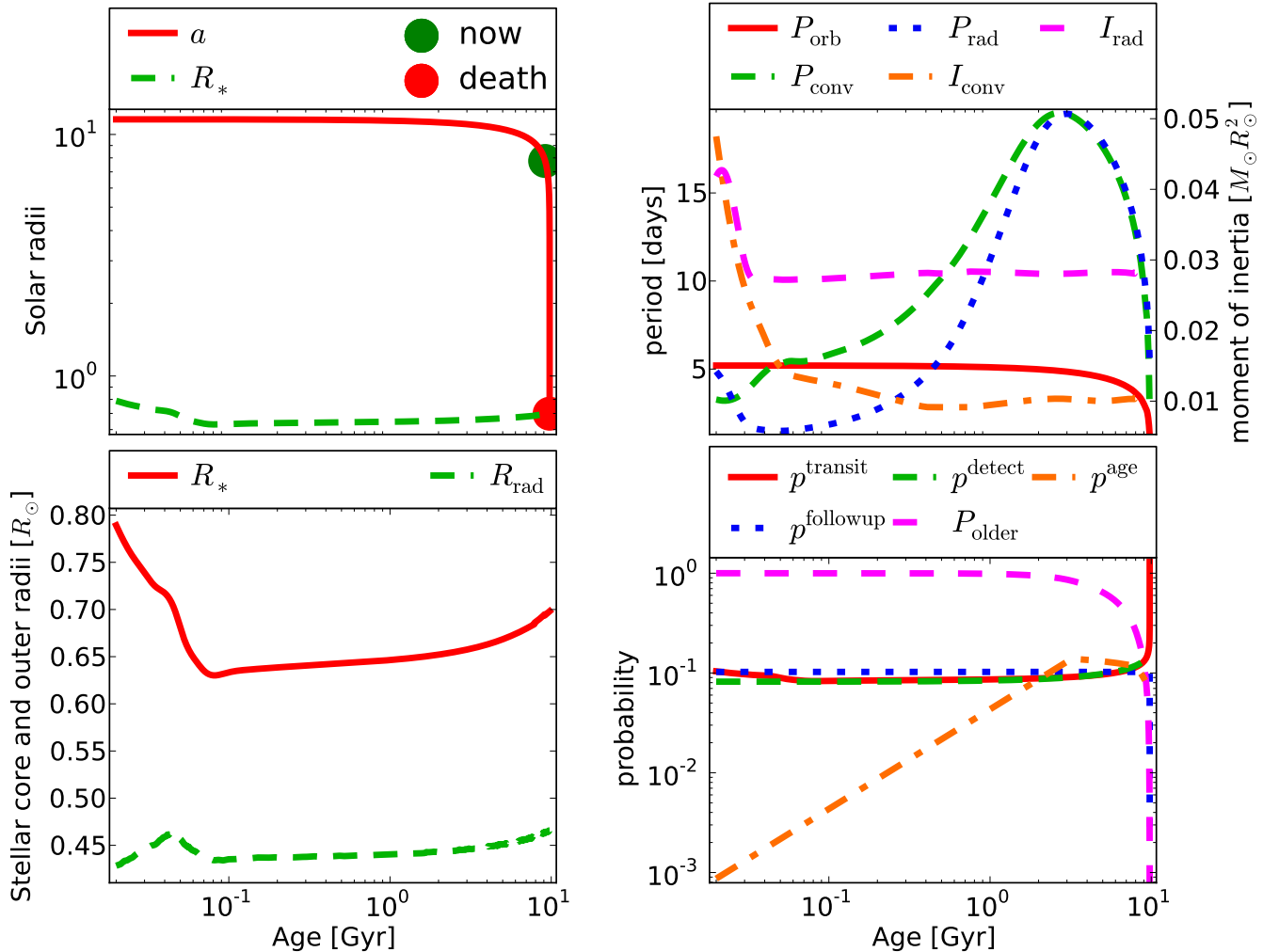}
		\caption{Example of the evolution of the HAT-P-20 system
		for $Q_*=10^6$. The various curves and points have the
		following meanings: $a$: the orbital semi-major axis in
		units of $R_\odot$; $R_*$: the radius of the star in units
		of $R_\odot$; $R_\mathrm{rad}$: the radius of the stellar
		core--envelope boundary in units of $R_\odot$;
		$P_\mathrm{orb}$: the orbital period in days;
		$P_\mathrm{conv}$, $P_\mathrm{rad}$: the spin
		periods in days of the convective and radiative zones of the
		star, respectively; $I_\mathrm{conv}$, $I_\mathrm{rad}$: the
		convective and radiative zone moments of inertia in solar
		units; and $p^\mathrm{transit}$, $p^\mathrm{detect}$,
		$p^\mathrm{followup}$, $p^\mathrm{age}$ and
		$P_\mathrm{older}$ are defined in Section \ref{sec: biases}.}
		\label{fig: example evolution}
	\end{center}
\end{figure*}

Figure \ref{fig: example evolution} shows an example of the evolution of
HAT-P-20 b, computed with $Q_*=10^6$. The left boundary of all our plots has
been placed at 30 Myr in order to show more details in the part of the
evolution that is important for this analysis. 

From the top right
plot one can see that for the first 55 Myr, due to its pre-main sequence
contraction, the star spins faster than the planet orbits. As a result the
semi-major axis increases (to a degree not noticeable on the plot), and the
star spins down, in spite of the fact that it is contracting. During
the subsequent approximately 35 Myr, the orbit and the convective zone of the
star are tidally locked. This lock is quickly lost due to the star 
shrinking (bottom left panel), causing the tidal coupling to sharply
decrease.

The core and the envelope of the star are clearly decoupled for the first
two Gyrs of the evolution. Initially, the core rotates slower than the
convective zone, due to the fact that its moment of inertia does not
change quite so much. However, at around 35 Myr, the situation is reversed
due to the stellar wind (and initially the planet) taking angular momentum
away from the convective zone, but not the core. Significant differential
rotation is maintained to an age of about 2 Gyr by the stellar wind,
which is gradually losing strength as the star spins down, until the core and
the envelope are completely coupled.

By around 3 Gyr, the orbit has shrunk enough for tidal torques to
once again dominate the rotational evolution of the star causing it to spin
up to a period of a few days as the planet inspirals.

\section{Observational Data}
\label{sec: list of systems}
From the more than one hundred currently known transiting planets, we based
our analysis on fifty--three. Those were selected to orbit stars with masses between
$0.25M_\odot$ and $1.25M_\odot$. The lower cutoff was imposed because we do
not have reliable stellar models for masses below this range. For masses
above the upper limit, the dissipation is likely dominated not by the
convective zone, which at this point is next to non-existent, but by some
dissipation mechanism in the radiative bulk of the star. Hence, assuming the
same $Q_*$ value applies beyond this point is not reasonable.

In particular, probably the planetary system most often given as an example
of a very fast tidal orbital evolution, WASP-18 b,
\citep[cf.]{Hellier_et_al_09, Hansen_10, Lai_12, Hellier_et_al_11a, 
Penev_Sasselov_11, Brown_et_al_11} is not among the planets included in this
work, due to the fact that its star lies above the $1.25M_\odot$ cutoff we
impose. The planetary orbit in this system, even just by itself, argues
strongly against efficient tidal dissipation in the star. However, as
discussed above, the mechanism of this dissipation is likely different than
for the majority of the exoplanet host stars found by transit searches.

Further, we restrict
our sample to only systems which are consistent with having a circular orbit,
and age limits quoted in the literature at least partially overlap with our
10 Gyr cutoff. Finally we exclude \textit{Kepler} and \textit{CoRoT} planets,
because they are generally subject to much different biases. Table \ref{tbl:
planets} lists the systems and their relevant parameters that were included
in our analysis, along with references to where those parameters were
published.

\begin{deluxetable*}{l|cccccccl}
	\tabletypesize{\small}
	\tablecaption{The Planetary Systems, their Parameters and the
	References used \label{tbl: planets}}
	\tablehead{
		\colhead{Planet Name} & 
		\colhead{$M_*$ ($M_\odot$)} &
		\colhead{$R_*$ ($R_\odot$)} & 
		\multicolumn{3}{c}{Stellar age (Gyr)} & 
		\colhead{Planet Mass} & 
		\colhead{Semi--major} & 
		\colhead{References}\\
		\colhead{} & \colhead{} & \colhead{} & \colhead{Nominal} &
		\colhead{Min} & \colhead{Max} & \colhead{($M_\mathrm{jup}$)} &
		\colhead{axis (AU)}& \colhead{}}
	\startdata
	HAT-P-3 b	& 0.917	& 0.799	& 1.6&	0.3&	4.5	& 0.591	& 0.03866 & 
		\citet{Chan_et_al_11}\\
	HAT-P-5 b	& 1.16	& 1.167	& 2.6&	0.8&	4.4	& 1.06	& 0.04075 &
		\citet{Bakos_et_al_07}\\
	HAT-P-10 b	& 0.83	& 0.79	& 7.9&	4.1&	11.7	& 0.487	& 0.0435 &
		\citet{Bakos_et_al_09a}\\
	HAT-P-12 b	& 0.733	& 0.701	& 2.5&	0.5&	4.5	& 0.211	& 0.0384 &
		\citet{Hartman_et_al_09b}\\
	HAT-P-13 b	& 1.22	& 1.56	& 5&	4.2&	7.5	& 0.851	& 0.0426 &
		\citet{Bakos_et_al_09b}\\
	&&&&&&&&\citet{Winn_et_al_10a}\\
	HAT-P-16 b	& 1.218	& 1.237	& 2&	1.2&	2.8	& 4.193	& 0.0413 &
		\citet{Buchhave_et_al_10}\\
	HAT-P-18 b	& 0.77	& 0.749	& 12.4&	6&	16.8	& 0.197	& 0.0559 &
		\citet{Hartman_et_al_11a}\\
	HAT-P-19 b	& 0.842	& 0.82	& 8.8&	3.6&	14	& 0.292	& 0.0466 &
		\citet{Hartman_et_al_11a}\\
	HAT-P-20 b	& 0.756	& 0.694	& 6.7&	2.9&	12.4	& 7.246	& 0.0361 &
		\citet{Bakos_et_al_11}\\
	HAT-P-22 b	& 0.916	& 1.04	& 12.4&	9.8&	15	& 2.147	& 0.0414 &
		\citet{Bakos_et_al_11}\\
	HAT-P-23 b	& 1.13	& 1.203	& 4&	3&	5	& 2.09	& 0.0232 &
		\citet{Bakos_et_al_11}\\
	HAT-P-24 b	& 1.191	& 1.317	& 2.8&	2.2&	3.4	& 0.685	& 0.0465 &
		\citet{Kipping_et_al_10}\\
	HAT-P-25 b	& 1.01	& 0.959	& 3.2&	0.9&	5.5	& 0.567	& 0.0466 &
		\citet{Quinn_et_al_12}\\
	HAT-P-26 b	& 0.816	& 0.788	& 9&	4.1&	12	& 0.059	& 0.0479 &
		\citet{Hartman_et_al_11b}\\
	HAT-P-27 b	& 0.945	& 0.898	& 4.4&	1.8&	8.2	& 0.66	& 0.0403 &
		\citet{Beky_et_al_11}\\
	HAT-P-28	& 1.025	& 1.103	& 6.1&	4.2&	8.7	& 0.626	& 0.0434 &
		\citet{Buchhave_et_al_11}\\
	HAT-P-29 b	& 1.207	& 1.224	& 2.2&	1.2&	3.2	& 0.778	& 0.0667 &
		\citet{Buchhave_et_al_11}\\
	HAT-P-32 b	& 1.16	& 1.219	& 2.7&	1.9&	3.5	& 0.86	& 0.0343 &
		\citet{Hartman_et_al_11c}\\
	HD 189733 b	& 0.82	& 0.73	& ?&	0.6&	?	& 1.13	& 0.03142 &
		\citet{Winn_et_al_06}\\
	&&&&&&&&\citet{Southworth_10}\\
	OGLE-TR-56 b	& 1.17	& 1.32	& 2.7&	2.6&	2.8	& 1.29	& 0.02386 &
		\citet{Pont_et_al_07}\\
	&&&&&&&&\citet{Southworth_10}\\
	OGLE-TR-113 b	& 0.78	& 0.77	& 0.7&	0.7&	10	& 1.32	& 0.0229 &
		\citet{Gillon_et_al_06}\\
	OGLE-TR-182 b	& 1.187	& 1.53	& 4.3&	2.4&	4.8	& 1.06	& 0.05205 &
		\citet{Southworth_10}\\
	Qatar-1 b	& 0.85	& 0.823	& 6&	6&	13	& 1.09	& 0.02343 &
		\citet{Alsubai_et_al_11}\\
	&&&&&&&&\citet{Sozzetti_04}\\
	&&&&&&&&\citet{Southworth_10}\\
	TrES-2		& 0.98	& 1	& 5.1&	2.4&	7.8	& 1.253	& 0.03556 &
		\citet{Sozzetti_et_al_07}\\
	&&&&&&&&\citet{Daemgen_et_al_09}\\
	&&&&&&&&\citet{Southworth_10}\\
	WASP-4 b	& 0.92	& 0.907	& 5.5&	3.5&	8.7	& 1.215	& 0.02312 &
		\citet{Wilson_et_al_08}\\
	&&&&&&&&\citet{Southworth_et_al_09a}\\
	&&&&&&&&\citet{Sanchis-Ojeda_et_al_11}\\
	WASP-5 b	& 1	& 1.084	& 3&	1.7&	4.4	& 1.637	& 0.02729 &
		\citet{Anderson_et_al_08}\\
	&&&&&&&&\citet{Southworth_et_al_09b}\\
	WASP-10 b	& 0.703	& 0.775	& 0.8&	0.6&	1	& 3.07	& 0.0371 &
		\citet{Christian_et_al_09}\\
	WASP-13 b	& 1.03	& 1.34	& 8.5&	3.6&	14	& 0.46	& 0.0527 &
		\citet{Skillen_et_al_09}\\
	WASP-16 b	& 1.022	& 0.946	& 2.3&	0.1&	8.1	& 0.855	& 0.0421 &
		\citet{Lister_et_al_09}\\
	WASP-19 b	& 0.97	& 0.99	& 5.5&	1&	14.5	& 1.168	& 0.01655 &
		\citet{Hellier_et_al_11a}\\
	WASP-21	b	& 1.01	& 1.06	& 12&	7&	17	& 0.3	& 0.052	 &
		\citet{Bouchy_et_al_10}\\
	WASP-22 b	& 1.1	& 1.13	& 3&	2&	4	& 0.56	& 0.0468 &
		\citet{Maxted_et_al_10}\\
	WASP-24 b	& 1.184	& 1.331	& 3.8&	2.5&	5.1	& 1.071	& 0.03651 &
		\citet{Street_et_al_10}\\
	WASP-25 b	& 1	& 0.92	& 0.02&	0.01&	3.98	& 0.58	& 0.0473 &
		\citet{Enoch_et_al_11a}\\
	WASP-26 b	& 1.12	& 1.34	& 6&	4&	8	& 1.02	& 0.04 &
		\citet{Smalley_et_al_10}\\
	WASP-28 b	& 1.08	& 1.05	& 5&	3&	8	& 0.91	& 0.0455 &
		\citet{West_et_al_10}\\
	WASP-34 b	& 1.01	& 0.93	& 6.7 &	2.2&	13.6	& 0.59	& 0.0524 &
		\citet{Smalley_et_al_11}\\
	WASP-35 b	& 1.07	& 1.09	& 5.01 & 3.85&	6.17	& 0.72	& 0.04317 &
		\citet{Enoch_et_al_11b}\\
	WASP-37 b	& 0.925	& 1.003	& 11&	7&	14	& 1.8	& 0.0446 &
		\citet{Simpson_et_al_11}\\
	WASP-38 b	& 1.203	& 1.331	& 5&	5&	14	& 2.691	& 0.07522 &
		\citet{Barros_et_al_11}\\
	WASP-39 b	& 0.93	& 0.895	& 9&	5&	12	& 0.28	& 0.0486 &
		\citet{Faedi_et_al_11}\\
	WASP-41 b	& 0.95	& 1.01	& 1.8 &	?&	?	& 0.92	& 0.04 &
		\citet{Maxted_et_al_11}\\
	WASP-43 b	& 0.58	& 0.598	& ? &	0.3&	?	& 1.78	& 0.0142 &
		\citet{Hellier_et_al_11b}\\
	WASP-44 b	& 0.951	& 0.927	& 0.9&	0.3&	1.9	& 0.889	& 0.03473 &
		\citet{Anderson_et_al_11}\\
	WASP-45 b	& 0.909	& 0.945	& 1.4&	0.4&	3.4	& 1.007	& 0.04054 &
		\citet{Anderson_et_al_11}\\
	WASP-46 b	& 0.956	& 0.917	& 1.4&	0.8&	1.8	& 2.101	& 0.02448 &
		\citet{Anderson_et_al_11}\\
	XO-2		& 0.98	& 0.97	& 5.3&	4.3&	6.3	& 0.57	& 0.0369  &
		\citet{Burke_et_al_07}\\
	55 Cnc e	& 0.905	& 0.943	& 10.2&	7.7&	12.7	& 0.027	& 0.0156 &
		\citet{Fischer_et_al_08}\\
	\enddata
\end{deluxetable*}

\section{Correcting for Observational Biases}
\label{sec: biases}

For each transiting planet system, we need to figure out what the probability
is that it would be observed and detected at any moment during its evolution.
We split this probability into four parts:

\begin{itemize}
	\item{$p^\mathrm{transit}$:} The geometric probability that the
		planet's transit is observed from the Earth.
	\item{$p^\mathrm{detect}$:} The probability that the orbital phase
		coverage of a survey allows the detection of the transit. 
	\item{$p^\mathrm{followup}$:} The probability that the transit
		candidate will be chosen for follow-up by the survey and
		confirmed.
	\item{$p^\mathrm{age}$:} The distribution of ages of target stars for
		transiting surveys expressed as a probability.
\end{itemize}

The final probability density that we use is the normalized product of these
four quantities.\\

The transit probability ($p^\mathrm{transit}$) is the simplest of the three
biases. It is simply proportional to the ratio of the stellar radius to the
orbital semi--major axis.\\

\begin{figure}
	\begin{center}
		\includegraphics[width=0.45\textwidth]{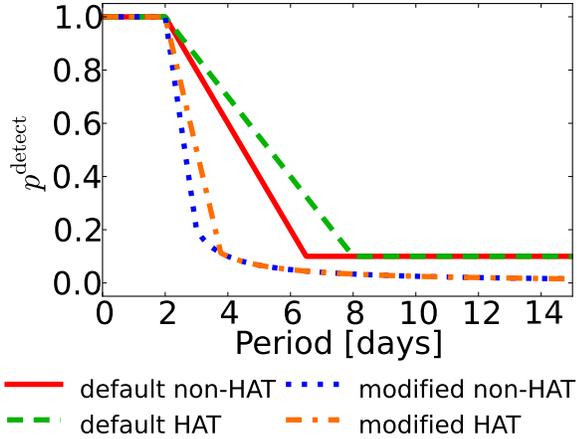}
		\caption{Assumed dependence of $p^\mathrm{detect}$ on
		orbital period for one station surveys (non-HAT) and the two
		station HAT survey.}
		\label{fig: Pdetect}
	\end{center}
\end{figure}

The detection probability ($p^\mathrm{detect}$) is a bit more complicated,
because the requirement for detecting a transit varies by system. For
example, to detect a relatively deep event around a quiet bright star,
observing only a few transits might be sufficient, while the detection of a
shallower transit around an active faint star might require many transits (we
incorporate the dependence of transit detection probability on stellar
activity into $p^\mathrm{followup}$). In addition, $p^\mathrm{detect}$ is not
a smooth function of orbital period, but rapidly oscillates, has sharp local
maxima or minima near periods close to an integer multiple of 24 hr, etc.
\citep[cf.][]{Collier_et_al_06, Smith_et_al_06, Burke_et_al_06,
Hartman_et_al_09a}. Instead of attempting to address all those complications,
we will assume a simple smooth dependence of $p^{detect}$ on the orbital
period and present results with two different prescriptions for $p^{detect}$.
The particular dependences of $p^{detect}$ on orbital period we use are given
in Figure \ref{fig: Pdetect}. These roughly follow the curves published by
various surveys for the recovery probability. \\

The long period tail of $p^{detect}$ is not well constrained,
and is survey dependent (hence the two different prescriptions). However, it
is also not particularly important, since planets with long periods are less
affected by tides, even for relatively small $Q_*$ values, and so our results
are not sensitive to this assumption.\\

The reason for prescribing shallower dependence of $p^\mathrm{detect}$ on
period for the HAT survey is that, unlike all other transit surveys, HAT
combines observations from two sites (one in Arizona and one in Hawaii) which
increases their sensitivity at longer periods.\\

In addition to  orbital period the detection probability
will depend on the brightness of the star and on the amplitude and frequency
dependence of stellar variability. However, since extrasolar planets are
typically found around main sequence stars, the stellar luminosity does not
vary much, and hence the stellar brightness dependence is mostly irrelevant for
our purposes, and is ignored in our model. The dependence on stellar
variability is generally complicated and difficult to quantify, so we include
it as part of the followup probability, and only in the general sense that
high stellar activity is associated with high stellar spin frequency and
hence the probability of a given transiting system being detected drops as
the stellar rotation period drops.\\

The follow-up probability ($p^\mathrm{followup}$) is the most difficult to
quantify, since it is subject to non-deterministic human evaluation and
limitations due to follow-up resources specific to each project. In this
work, we will idealize the situation and assume that it only depends on the
rotational period of the star ($P_\mathrm{rot}$), or its projected equatorial
rotation velocity $v^*\sin i$ ($v^*$ is the the equatorial rotation velocity
of the star and $i$ is the angle between stellar rotational axis and the line
of sight).  Stars rotating fast have a smaller probability to be chosen for
follow-up because the radial velocity precision will be limited by the
rotational broadening of the spectral lines.  In addition, a star's
variability and hence a survey's ability to recognize a transit signal are
correlated with the stellar rotation, so there is a bias against detecting
transits for quickly rotating stars, which we did not include in
$p^\mathrm{detect}$. In this work, we test the dependence of our results on
$p^\mathrm{followup}$ by assuming two different forms for it: (1) constant
and (2) constant up to $v^*\sin i<20$ km s$^{-1}$, followed by an exponential
decay with the follow-up probability reaching half its maximal value at
$v^*\sin i=40$ km s$^{-1}$.\\

\begin{figure}
	\begin{center}
		\includegraphics[width=0.45\textwidth]{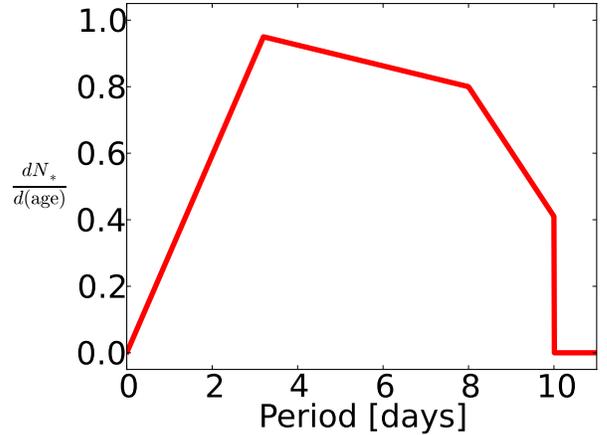}
		\caption{Assumed distribution of ages for the target stars
		of transit search projects.}
		\label{fig: Page}
	\end{center}
\end{figure}

Finally, our prescription for $p^\mathrm{age}$ is based on the dotted line of
Figure 3 of \citet{Takeda_et_al_07}. In particular we use a piecewise linear
approximation to their curve (see Figure \ref{fig: Page}). In addition we
also present results with $p^\mathrm{age}=\mathrm{const}$.\\

In the right panel of Figure \ref{fig: example evolution}, we show the
computed probability density functions $p^\mathrm{transit}$,
$p^\mathrm{detect}$, $p^\mathrm{followup}$ and $p^\mathrm{age}$ that
correspond to the evolution of for HAT-P-20 b presented in the other panels
of the same figure. We also show the resulting $P_\mathrm{older}$, computed
as:
\begin{equation}
	P_\mathrm{older}(\mathrm{age})=\frac{
		\int_\mathrm{age}^\mathrm{death} p^\mathrm{transit}
		p^\mathrm{detect} p^\mathrm{followup} p^\mathrm{age}
		d(\mathrm{age})
	}{
		\int_0^\mathrm{death} p^\mathrm{transit} p^\mathrm{detect}
		p^\mathrm{followup} p^\mathrm{age} d(\mathrm{age})
	}
	\label{eq: Polder}
\end{equation}

In other words, $P_\mathrm{older}$ represents the probability to observe a
planet at its current age or older. As a planet's orbit decays, we have less
and less chance to observe the planet before it is tidally disrupted.

\section{Testing $Q_*$ values}
\label{sec: testing Q*}
In this section, we outline the procedure we use to determine if a given value
of the $Q_*$ parameter is consistent with the observed set of exoplanet
systems.\\

We begin by finding an initial (5 Myr after stellar birth) semi--major axis
for each planetary system, which after following the orbital evolution, 
according to Equation (\ref{eq: adot}---\ref{eq: orbital evol last}),
to the present time, results in the observed value of the semi--major axis.\\

We then continue the evolution until one of the following happens. 
\begin{enumerate}
	\item The semi--major axis of the orbit falls below (1) the radius of
		the star or (2) the Roche radius for the tidal destruction of
		the planet, whichever comes first. If only tidal decay drives
		the orbital evolution, we expect planets to spend very little
		time between their Roche limit and the stellar surface (when
		the former is larger), and so for our purposes the distinction
		between the stellar surface and Roche radius is unimportant:
		planets that have crossed their Roche radii are as good as
		gone. \label{cond: planet death}
	\item The star reaches the end of its main--sequence lifetime.
		\label{cond: star death}
	\item The system reaches an age of $T_\mathrm{max}=10$ Gyr and
		neither of the above conditions has occurred. \label{cond:
		max age}
\end{enumerate}

Having the complete time evolution of each system, we use
$p^\mathrm{transit}$, $p^\mathrm{detect}$, $p^\mathrm{followup}$, and
$p^\mathrm{age}$ from Section \ref{sec: biases} to calculate the probability
that a random observation throughout its lifetime will catch it at any given
moment. Integrating this probability density from the present age of the
system onward gives us the probability of observing this system no earlier
than its present age -- $P_\mathrm{older}$.  Hence, by definition, we expect
$P_\mathrm{older}$ to be uniformly distributed in the range $(0,1)$. If we
find it is not, this means that either we did not properly account for some
observational or astrophysical bias or the orbital evolution computed is not
correct. Assuming the various biases are handled correctly, and the orbital
evolution equations (Equations (\ref{eq: adot}) -- (\ref{eq: deltaL})) are
appropriate, departures from uniformity will be due to a mismatch between the
assumed $Q_*$ and the real one. So a KS test allows us to reject values of
$Q_*$ which are inconsistent with the currently observed population of
extrasolar planets, around stars for which the dissipation is likely
dominated by the convective zone.\\

\begin{figure}
	\begin{center}
		\includegraphics[width=0.45\textwidth]{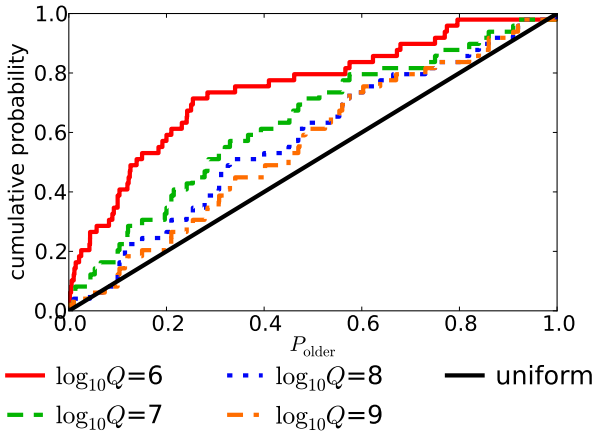}
		\caption{Cumulative distribution function of
		$P_\mathrm{older}$ for various values of $Q_*$.}
		\label{fig: CDF(Polder)}
	\end{center}
\end{figure}

\begin{figure}
	\begin{center}
		\includegraphics[width=0.45\textwidth]{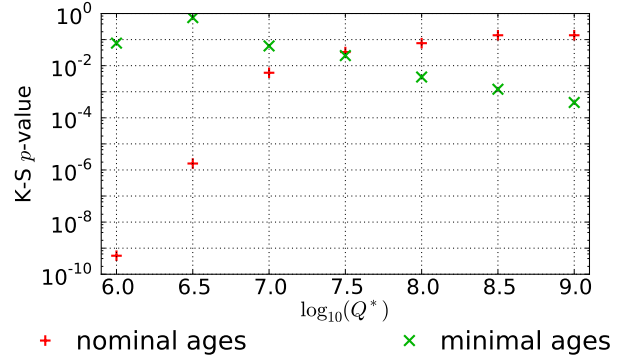}
		\caption{KS test $p$-values of the comparison of the
		cumulative distributions of $P_\mathrm{older}$, calculated by
		assuming that all stars' ages are equal to the nominal
		values (red +) and the minimum allowed value (blue x), 
		against a uniform distribution for various values of $Q_*$.}
		\label{fig: NominalAgeKSp}
	\end{center}
\end{figure}

A plot of the calculated cumulative distribution function of
$P_\mathrm{older}$ ($\mathrm{CDF}(P_\mathrm{older})$) appropriately corrected
for observational biases and assuming the nominal values for the observed
system parameters quoted in the literature for various values of $Q_*$ is
presented in Figure  \ref{fig: CDF(Polder)}. The KS test $p$-values
corresponding to comparing those curves against a uniform distribution are
shown as the red (+) symbols in Figure \ref{fig: NominalAgeKSp}.\\

There are several observational uncertainties that may affect our results.
However, the uncertainty in the stellar ages (often a factor of several)
dominates over the uncertainties of all other quantities. Thus, this is the
only uncertainty we will account for. To demonstrate the possible impact on
our conclusions, we compute cumulative distributions of $P_\mathrm{older}$
with two assumptions for the actual ages of stars --- the nominal ages and
the lower end of the age range given by the corresponding publication. The
resulting $p$-values obtained by performing a KS test against a uniform
distribution are presented in Figure \ref{fig: NominalAgeKSp}. 

The shift to smaller $Q_*$ values when the exoplanetary systems are assumed
systematically younger makes sense. A smaller present age of the system means
that the interval between now and when one of the terminal conditions
described at the beginning of this section occurs represents a larger
fraction of the total lifetime of the system. To offset this, a smaller value
of $Q_*$ is needed, shortening the future life of the system.

Because the uncertainty is different for each system and because it depends
on factors like the method for determining stellar ages, follow up
instrumentation, stellar models used, etc., we cannot modify the expected
distribution to include such an uncertainty. Instead one can show that if we
prescribe a distribution for the age of each system, the statistical
$p$-value corresponding to a given $Q_*$ is the expectation of the usual KS
$p$-value over the distribution of ages. Since evaluating this expectation
requires taking an integral over a space which has as many dimensions as the
observed transiting planets, we use a Monte-Carlo approach to calculate it.
The precise procedure is as follows.
\begin{enumerate}
	\item Calculate the evolution of each system for a set of present
		ages covering the observationally allowed range.
	\item Draw a random age for each system from some prescribed
		distribution.
	\item Calculate the corresponding $P_\mathrm{older}$ and perform a KS
		test.
	\item Average the results of many such iterations to get the final
		$p$-value, which incorporates the uncertainty in the system
		ages.
\end{enumerate}

Since we do not know the appropriate distribution to assume for each system
we will consider two options: a uniform distribution over the allowed
interval of ages, and a normal distribution truncated at the age limits
centered on the nominal age, with a standard deviation equal to one quarter
of the given age range.

\section{Results}
\label{sec: results}

\begin{deluxetable*}{l|ccccc}[!t]
	\tabletypesize{\small}
	\tablecaption{The Various Assumptions used when Performing the K--S
	Tests for the Various $Q_*$ Values. \label{tbl: assumptions}}
	\tablehead{
		\colhead{Label} & 
		\colhead{$p^\mathrm{detect}$} & 
		\colhead{$p^\mathrm{followup}$} & 
		\colhead{$p^\mathrm{age}$} & 
		\colhead{Age Distribution} &
		\colhead{$\tau_c (\mathrm{Myr})$}}
	\startdata
		Default 
		& Red (solid) curve in Fig. \ref{fig: Pdetect}
	 	& Decreasing for $v^*\sin i>20$ km/s
	 	& As plotted in Fig. \ref{fig: Page}
	 	& Uniform
		& 5\\
		Mod $p^\mathrm{detect}$
		& Blue (dashed) curve in Fig. \ref{fig: Pdetect}
	 	& Decreasing for $v^*\sin i>20$ km/s
	 	& As plotted in Fig. \ref{fig: Page}
	 	& Uniform
		& 5\\
		Uniform $p^\mathrm{followup}$
		& Red (solid) curve in Fig. \ref{fig: Pdetect}
		& Uniform
		& As plotted in Fig. \ref{fig: Page}
		& Uniform
		& 5\\
		Uniform $p^\mathrm{age}$
		& Red (solid) curve in Fig. \ref{fig: Pdetect}
	 	& Decreasing for $v^*\sin i>20$ km/s
	 	& As plotted in Fig. \ref{fig: Page}
	 	& Uniform
		& 5\\
		Normal age dist.
		& Red (solid) curve in Fig. \ref{fig: Pdetect}
	 	& Decreasing for $v^*\sin i>20$ km/s
	 	& As plotted in Fig. \ref{fig: Page}
	 	& Normal
		& 5\\
		Coupled 
		& Red (solid) curve in Fig. \ref{fig: Pdetect}
	 	& Decreasing for $v^*\sin i>20$ km/s
	 	& As plotted in Fig. \ref{fig: Page}
	 	& Uniform
		& 0\\
	\enddata
\end{deluxetable*}

\begin{figure}
	\begin{center}
		\includegraphics[width=0.45\textwidth]{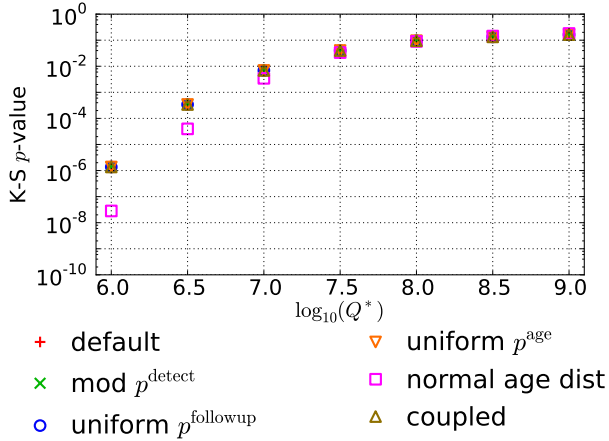}
		\caption{KS test $p$-values as a function of $Q_*$ for the
		range of assumptions detailed in Table \ref{tbl:
		assumptions}.}
		\label{fig: AgeSampledKSp}
	\end{center}
\end{figure}

The procedure described in the previous section was performed six times with
various assumptions for $p^\mathrm{detect}$, $p^\mathrm{followup}$ and
$p^\mathrm{age}$ as discussed in Section \ref{sec: biases}, for the two per
system age distributions described above and finally we considered a case
where we do not allow the convective and radiative zones to rotate at
different frequencies (i.e., $\tau_c=0$). Table \ref{tbl: assumptions} lists
the set of assumptions we consider, and the resulting KS test $p$-values are
plotted in Figure \ref{fig: AgeSampledKSp}.\\

Evidently, only the assumption about how stellar ages are distributed
over their observationally allowed ranges makes a noticeable difference.
This is not altogether surprising considering that of all the assumed
probability distributions this is the most poorly constrained observationally. 
The core--envelope coupling timescale also has little effect, since allowing
differential rotation or not never results in the star rotating faster than
the planet except during a very short period during the pre-main sequence
phase (like in Figure \ref{fig: example evolution}).

All assumptions considered lead to very similar conclusions: at the 1\%
level $\log_{10}Q_*>7$. The different assumptions, fortunately, lead to
differences only in the low $p$-value range of $Q_*$.

\section{Discussion}
\label{sec: discussion}
The constraints on the stellar dissipation parameter $Q_*$ we derive based on
the transiting planet systems detected by ground based transit surveys
($Q_*>10^7$) are inconsistent with constraints derived from observing the
circularization of binary stellar systems in open clusters. 

\citet{Zahn_89} shows that, in order for tides to suppress the eccentricity in
binary stars up to the observed circularization cutoff period,
$Q_*\sim10^7$ is required during the pre-main sequence phase, which
corresponds to a $p$-value of 1\% in our analysis. One can somewhat circumvent
this marginal contradiction by assuming an evolution of $Q_*$ as the stellar
structure changes. However, \citet{Meibom_Mathieu_05} find that in order to
explain the observed rate of circularization during the main sequence phase
even smaller values are required---$Q_*\sim10^5$.

One way to reconcile this apparent inconsistency is to note that the ratio of
tidal frequency to stellar rotation frequency is very different for binary
star circularization and tidal inspiral of a planet onto its star. For binary
stars, the two components of the system are synchronized on a very short time
scale (compared to circularization), so the tidal frequency is exactly twice
the stellar rotational frequency. In the case of an exoplanet inspiral the
tidal frequency is much higher than the stellar rotation.\\

This different frequency might lead to two different mechanisms dominating
the dissipation in the two cases. \citet{Ogilvie_Lin_04, Ogilvie_Lin_07} and 
\citet{Wu_05a, Wu_05b} point out that if the tidal frequency is within a
factor of two of the stellar rotational frequency, inertial modes are
resonantly excited in the star, which could lead to strongly enhanced shear
and hence dissipation. Since the inertial mode frequencies are restricted
to lie between $-2\omega_*$ and $2\omega_*$, this mechanism cannot operate in
the case of exoplanet systems. The currently favored mechanism of dissipation
for low mass stars in the frequency regime of exoplanetary tides is turbulent
dissipation \citep{Zahn_66, Zahn_89, Goldreich_Nicholson_77,
Goldreich_Keeley_77, Goodman_Oh_97, Penev_Sasselov_Robinson_Demarque_07,
Penev_Sasselov_Robinson_Demarque_09, Penev_Barranco_Sasselov_09,
Penev_Barranco_Sasselov_11, Penev_Sasselov_11}. This less efficient mechanism
could result in dissipation efficiencies consistent with the constraints
derived in this paper. In particular \citet{Penev_Sasselov_11}, based on
direct simulations of turbulent dissipation find
$Q_*\sim\mathrm{few}\times10^8$ to $10^9$, consistent with our results here.

Finally, \citet{Schlaufman_Lin_Ida_10} argue that the sample of Kepler
planets favors $10^6<Q_*<10^7$, apparently outside our range. However, their
analysis does not even consider $Q_*$ values above $10^7$ (other than
infinity), and they do not derive the statistical significance of their
limits, or the sensitivity of their result on the various assumptions
included in their model (e.g., the conversion of mass to radius and the
exoplanet population synthesis models they use). All this makes it difficult
to make firm statements about the (in)consistency of the two results. The
best way to address this would be to repeat our analysis for the sample of
Kepler planets, but this is clearly outside the scope of this article.

Our model suggests that the earliest stages of a close-in planet's dynamical
history may be more complicated than widely considered. The top right panel of
Figure \ref{fig: example evolution} shows that, early on, HAT-P-20 rotated
more quickly than its planet revolved and only after 35 Myr did the
situation reverse. Consequently, the tidal torque exerted by the star
switched signs at this point. If, for example, the planet were brought
close-in through gas disk migration during its first tens of Myrs, then
presumably there was a competition between torques from the gas disk driving
the planet in and tidal torques from the star driving the planet out.
\citet{Lin_et_al_96} pointed out the role that such tidal torques might play
in stopping 51 Peg b's inward gas disk migration but favored clearing of the
gas disk very near the star for halting the inward migration. With so many
more exoplanets in our sample now, many with much shorter orbital periods
than 51 Peg b, we plan to revisit this topic.

\bibliography{bibliography}

\begin{thebibliography}{125}
\expandafter\ifx\csname natexlab\endcsname\relax\def\natexlab#1{#1}\fi

\bibitem[{{Allain}(1998)}]{Allain_98}
{Allain}, S. 1998, \aap, 333, 629

\bibitem[{{Alsubai} {et~al.}(2011){Alsubai}, {Parley}, {Bramich}, {West},
  {Sorensen}, {Collier Cameron}, {Latham}, {Horne}, {Anderson}, {Bakos},
  {Brown}, {Buchhave}, {Esquerdo}, {Everett}, {F{\.z}r{\'e}sz}, {Hartman},
  {Hellier}, {Miller}, {Pollacco}, {Quinn}, {Smith}, {Stefanik}, \&
  {Szentgyorgyi}}]{Alsubai_et_al_11}
{Alsubai}, K.~A., {Parley}, N.~R., {Bramich}, D.~M., {West}, R.~G., {Sorensen},
  P.~M., {Collier Cameron}, A., {Latham}, D.~W., {Horne}, K., {Anderson},
  D.~R., {Bakos}, G.~{\'A}., {Brown}, D.~J.~A., {Buchhave}, L.~A., {Esquerdo},
  G.~A., {Everett}, M.~E., {F{\.z}r{\'e}sz}, G., {Hartman}, J.~D., {Hellier},
  C., {Miller}, G.~M., {Pollacco}, D., {Quinn}, S.~N., {Smith}, J.~C.,
  {Stefanik}, R.~P., \& {Szentgyorgyi}, A. 2011, \mnras, 417, 709

\bibitem[{{Anderson} {et~al.}(2011){Anderson}, {Collier Cameron}, {Gillon},
  {Hellier}, {Jehin}, {Lendl}, {Maxted}, {Queloz}, {Smalley}, {Smith},
  {Triaud}, {West}, {Pepe}, {Pollacco}, {S{\'e}gransan}, {Todd}, \&
  {Udry}}]{Anderson_et_al_11}
{Anderson}, D.~R., {Collier Cameron}, A., {Gillon}, M., {Hellier}, C., {Jehin},
  E., {Lendl}, M., {Maxted}, P.~F.~L., {Queloz}, D., {Smalley}, B., {Smith},
  A.~M.~S., {Triaud}, A.~H.~M.~J., {West}, R.~G., {Pepe}, F., {Pollacco}, D.,
  {S{\'e}gransan}, D., {Todd}, I., \& {Udry}, S. 2011, ArXiv e-prints

\bibitem[{{Anderson} {et~al.}(2008){Anderson}, {Gillon}, {Hellier}, {Maxted},
  {Pepe}, {Queloz}, {Wilson}, {Collier Cameron}, {Smalley}, {Lister},
  {Bentley}, {Blecha}, {Christian}, {Enoch}, {Hebb}, {Horne}, {Irwin}, {Joshi},
  {Kane}, {Marmier}, {Mayor}, {Parley}, {Pollacco}, {Pont}, {Ryans},
  {S{\'e}gransan}, {Skillen}, {Street}, {Udry}, {West}, \&
  {Wheatley}}]{Anderson_et_al_08}
{Anderson}, D.~R., {Gillon}, M., {Hellier}, C., {Maxted}, P.~F.~L., {Pepe}, F.,
  {Queloz}, D., {Wilson}, D.~M., {Collier Cameron}, A., {Smalley}, B.,
  {Lister}, T.~A., {Bentley}, S.~J., {Blecha}, A., {Christian}, D.~J., {Enoch},
  B., {Hebb}, L., {Horne}, K., {Irwin}, J., {Joshi}, Y.~C., {Kane}, S.~R.,
  {Marmier}, M., {Mayor}, M., {Parley}, N.~R., {Pollacco}, D.~L., {Pont}, F.,
  {Ryans}, R., {S{\'e}gransan}, D., {Skillen}, I., {Street}, R.~A., {Udry}, S.,
  {West}, R.~G., \& {Wheatley}, P.~J. 2008, \mnras, 387, L4

\bibitem[{{Bakos} {et~al.}(2011){Bakos}, {Hartman}, {Torres}, {Latham},
  {Kov{\'a}cs}, {Noyes}, {Fischer}, {Johnson}, {Marcy}, {Howard}, {Kipping},
  {Esquerdo}, {Shporer}, {B{\'e}ky}, {Buchhave}, {Perumpilly}, {Everett},
  {Sasselov}, {Stefanik}, {L{\'a}z{\'a}r}, {Papp}, \&
  {S{\'a}ri}}]{Bakos_et_al_11}
{Bakos}, G.~{\'A}., {Hartman}, J., {Torres}, G., {Latham}, D.~W., {Kov{\'a}cs},
  G., {Noyes}, R.~W., {Fischer}, D.~A., {Johnson}, J.~A., {Marcy}, G.~W.,
  {Howard}, A.~W., {Kipping}, D., {Esquerdo}, G.~A., {Shporer}, A., {B{\'e}ky},
  B., {Buchhave}, L.~A., {Perumpilly}, G., {Everett}, M., {Sasselov}, D.~D.,
  {Stefanik}, R.~P., {L{\'a}z{\'a}r}, J., {Papp}, I., \& {S{\'a}ri}, P. 2011,
  \apj, 742, 116

\bibitem[{{Bakos} {et~al.}(2009{\natexlab{a}}){Bakos}, {Howard}, {Noyes},
  {Hartman}, {Torres}, {Kov{\'a}cs}, {Fischer}, {Latham}, {Johnson}, {Marcy},
  {Sasselov}, {Stefanik}, {Sip{\H o}cz}, {Kov{\'a}cs}, {Esquerdo}, {P{\'a}l},
  {L{\'a}z{\'a}r}, {Papp}, \& {S{\'a}ri}}]{Bakos_et_al_09b}
{Bakos}, G.~{\'A}., {Howard}, A.~W., {Noyes}, R.~W., {Hartman}, J., {Torres},
  G., {Kov{\'a}cs}, G., {Fischer}, D.~A., {Latham}, D.~W., {Johnson}, J.~A.,
  {Marcy}, G.~W., {Sasselov}, D.~D., {Stefanik}, R.~P., {Sip{\H o}cz}, B.,
  {Kov{\'a}cs}, G., {Esquerdo}, G.~A., {P{\'a}l}, A., {L{\'a}z{\'a}r}, J.,
  {Papp}, I., \& {S{\'a}ri}, P. 2009{\natexlab{a}}, \apj, 707, 446

\bibitem[{{Bakos} {et~al.}(2009{\natexlab{b}}){Bakos}, {P{\'a}l}, {Torres},
  {Sip{\H o}cz}, {Latham}, {Noyes}, {Kov{\'a}cs}, {Hartman}, {Esquerdo},
  {Fischer}, {Johnson}, {Marcy}, {Butler}, {Howard}, {Sasselov}, {Kov{\'a}cs},
  {Stefanik}, {L{\'a}z{\'a}r}, {Papp}, \& {S{\'a}ri}}]{Bakos_et_al_09a}
{Bakos}, G.~{\'A}., {P{\'a}l}, A., {Torres}, G., {Sip{\H o}cz}, B., {Latham},
  D.~W., {Noyes}, R.~W., {Kov{\'a}cs}, G., {Hartman}, J., {Esquerdo}, G.~A.,
  {Fischer}, D.~A., {Johnson}, J.~A., {Marcy}, G.~W., {Butler}, R.~P.,
  {Howard}, A.~W., {Sasselov}, D.~D., {Kov{\'a}cs}, G., {Stefanik}, R.~P.,
  {L{\'a}z{\'a}r}, J., {Papp}, I., \& {S{\'a}ri}, P. 2009{\natexlab{b}}, \apj,
  696, 1950

\bibitem[{{Bakos} {et~al.}(2007){Bakos}, {Shporer}, {P{\'a}l}, {Torres},
  {Kov{\'a}cs}, {Latham}, {Mazeh}, {Ofir}, {Noyes}, {Sasselov}, {Bouchy},
  {Pont}, {Queloz}, {Udry}, {Esquerdo}, {Sip{\H o}cz}, {Kov{\'a}cs},
  {Stefanik}, {L{\'a}z{\'a}r}, {Papp}, \& {S{\'a}ri}}]{Bakos_et_al_07}
{Bakos}, G.~{\'A}., {Shporer}, A., {P{\'a}l}, A., {Torres}, G., {Kov{\'a}cs},
  G., {Latham}, D.~W., {Mazeh}, T., {Ofir}, A., {Noyes}, R.~W., {Sasselov},
  D.~D., {Bouchy}, F., {Pont}, F., {Queloz}, D., {Udry}, S., {Esquerdo}, G.,
  {Sip{\H o}cz}, B., {Kov{\'a}cs}, G., {Stefanik}, R., {L{\'a}z{\'a}r}, J.,
  {Papp}, I., \& {S{\'a}ri}, P. 2007, \apjl, 671, L173

\bibitem[{{Barnes} \& {Sofia}(1996)}]{Barnes_Sofia_96}
{Barnes}, S. \& {Sofia}, S. 1996, \apj, 462, 746

\bibitem[{{Barros} {et~al.}(2011){Barros}, {Faedi}, {Collier Cameron},
  {Lister}, {McCormac}, {Pollacco}, {Simpson}, {Smalley}, {Street}, {Todd},
  {Triaud}, {Boisse}, {Bouchy}, {H{\'e}brard}, {Moutou}, {Pepe}, {Queloz},
  {Santerne}, {Segransan}, {Udry}, {Bento}, {Butters}, {Enoch}, {Haswell},
  {Hellier}, {Keenan}, {Miller}, {Moulds}, {Norton}, {Parley}, {Skillen},
  {Watson}, {West}, \& {Wheatley}}]{Barros_et_al_11}
{Barros}, S.~C.~C., {Faedi}, F., {Collier Cameron}, A., {Lister}, T.~A.,
  {McCormac}, J., {Pollacco}, D., {Simpson}, E.~K., {Smalley}, B., {Street},
  R.~A., {Todd}, I., {Triaud}, A.~H.~M.~J., {Boisse}, I., {Bouchy}, F.,
  {H{\'e}brard}, G., {Moutou}, C., {Pepe}, F., {Queloz}, D., {Santerne}, A.,
  {Segransan}, D., {Udry}, S., {Bento}, J., {Butters}, O.~W., {Enoch}, B.,
  {Haswell}, C.~A., {Hellier}, C., {Keenan}, F.~P., {Miller}, G.~R.~M.,
  {Moulds}, V., {Norton}, A.~J., {Parley}, N., {Skillen}, I., {Watson}, C.~A.,
  {West}, R.~G., \& {Wheatley}, P.~J. 2011, \aap, 525, A54+

\bibitem[{{Bayliss} {et~al.}(2010){Bayliss}, {Winn}, {Mardling}, \&
  {Sackett}}]{Bayliss_et_al_10}
{Bayliss}, D.~D.~R., {Winn}, J.~N., {Mardling}, R.~A., \& {Sackett}, P.~D.
  2010, \apjl, 722, L224

\bibitem[{{B{\'e}ky} {et~al.}(2011){B{\'e}ky}, {Bakos}, {Hartman}, {Torres},
  {Latham}, {Jord{\'a}n}, {Arriagada}, {Bayliss}, {Kiss}, {Kov{\'a}cs},
  {Quinn}, {Marcy}, {Howard}, {Fischer}, {Johnson}, {Esquerdo}, {Noyes},
  {Buchhave}, {Sasselov}, {Stefanik}, {Perumpilly}, {L{\'a}z{\'a}r}, {Papp}, \&
  {S{\'a}ri}}]{Beky_et_al_11}
{B{\'e}ky}, B., {Bakos}, G.~{\'A}., {Hartman}, J., {Torres}, G., {Latham},
  D.~W., {Jord{\'a}n}, A., {Arriagada}, P., {Bayliss}, D., {Kiss}, L.~L.,
  {Kov{\'a}cs}, G., {Quinn}, S.~N., {Marcy}, G.~W., {Howard}, A.~W., {Fischer},
  D.~A., {Johnson}, J.~A., {Esquerdo}, G.~A., {Noyes}, R.~W., {Buchhave},
  L.~A., {Sasselov}, D.~D., {Stefanik}, R.~P., {Perumpilly}, G.,
  {L{\'a}z{\'a}r}, J., {Papp}, I., \& {S{\'a}ri}, P. 2011, \apj, 734, 109

\bibitem[{{Bouchy} {et~al.}(2010){Bouchy}, {Hebb}, {Skillen}, {Collier
  Cameron}, {Smalley}, {Udry}, {Anderson}, {Boisse}, {Enoch}, {Haswell},
  {H{\'e}brard}, {Hellier}, {Joshi}, {Kane}, {Maxted}, {Mayor}, {Moutou},
  {Pepe}, {Pollacco}, {Queloz}, {S{\'e}gransan}, {Simpson}, {Smith},
  {Stempels}, {Street}, {Triaud}, {West}, \& {Wheatley}}]{Bouchy_et_al_10}
{Bouchy}, F., {Hebb}, L., {Skillen}, I., {Collier Cameron}, A., {Smalley}, B.,
  {Udry}, S., {Anderson}, D.~R., {Boisse}, I., {Enoch}, B., {Haswell}, C.~A.,
  {H{\'e}brard}, G., {Hellier}, C., {Joshi}, Y., {Kane}, S.~R., {Maxted},
  P.~F.~L., {Mayor}, M., {Moutou}, C., {Pepe}, F., {Pollacco}, D., {Queloz},
  D., {S{\'e}gransan}, D., {Simpson}, E.~K., {Smith}, A.~M.~S., {Stempels},
  H.~C., {Street}, R., {Triaud}, A.~H.~M.~J., {West}, R.~G., \& {Wheatley},
  P.~J. 2010, \aap, 519, A98+

\bibitem[{{Bouwman} {et~al.}(2006){Bouwman}, {Lawson}, {Dominik}, {Feigelson},
  {Henning}, {Tielens}, \& {Waters}}]{Bouwman_et_al_06}
{Bouwman}, J., {Lawson}, W.~A., {Dominik}, C., {Feigelson}, E.~D., {Henning},
  T., {Tielens}, A.~G.~G.~M., \& {Waters}, L.~B.~F.~M. 2006, \apjl, 653, L57

\bibitem[{{Brown} {et~al.}(2011){Brown}, {Collier Cameron}, {Hall}, {Hebb}, \&
  {Smalley}}]{Brown_et_al_11}
{Brown}, D.~J.~A., {Collier Cameron}, A., {Hall}, C., {Hebb}, L., \& {Smalley},
  B. 2011, \mnras, 415, 605

\bibitem[{{Buchhave} {et~al.}(2010){Buchhave}, {Bakos}, {Hartman}, {Torres},
  {Kov{\'a}cs}, {Latham}, {Noyes}, {Esquerdo}, {Everett}, {Howard}, {Marcy},
  {Fischer}, {Johnson}, {Andersen}, {F{\H u}r{\'e}sz}, {Perumpilly},
  {Sasselov}, {Stefanik}, {B{\'e}ky}, {L{\'a}z{\'a}r}, {Papp}, \&
  {S{\'a}ri}}]{Buchhave_et_al_10}
{Buchhave}, L.~A., {Bakos}, G.~{\'A}., {Hartman}, J.~D., {Torres}, G.,
  {Kov{\'a}cs}, G., {Latham}, D.~W., {Noyes}, R.~W., {Esquerdo}, G.~A.,
  {Everett}, M., {Howard}, A.~W., {Marcy}, G.~W., {Fischer}, D.~A., {Johnson},
  J.~A., {Andersen}, J., {F{\H u}r{\'e}sz}, G., {Perumpilly}, G., {Sasselov},
  D.~D., {Stefanik}, R.~P., {B{\'e}ky}, B., {L{\'a}z{\'a}r}, J., {Papp}, I., \&
  {S{\'a}ri}, P. 2010, \apj, 720, 1118

\bibitem[{{Buchhave} {et~al.}(2011){Buchhave}, {Bakos}, {Hartman}, {Torres},
  {Latham}, {Andersen}, {Kov{\'a}cs}, {Noyes}, {Shporer}, {Esquerdo},
  {Fischer}, {Johnson}, {Marcy}, {Howard}, {B{\'e}ky}, {Sasselov}, {F{\H
  u}r{\'e}sz}, {Quinn}, {Stefanik}, {Szklen{\'a}r}, {Berlind}, {Calkins},
  {L{\'a}z{\'a}r}, {Papp}, \& {S{\'a}ri}}]{Buchhave_et_al_11}
{Buchhave}, L.~A., {Bakos}, G.~{\'A}., {Hartman}, J.~D., {Torres}, G.,
  {Latham}, D.~W., {Andersen}, J., {Kov{\'a}cs}, G., {Noyes}, R.~W., {Shporer},
  A., {Esquerdo}, G.~A., {Fischer}, D.~A., {Johnson}, J.~A., {Marcy}, G.~W.,
  {Howard}, A.~W., {B{\'e}ky}, B., {Sasselov}, D.~D., {F{\H u}r{\'e}sz}, G.,
  {Quinn}, S.~N., {Stefanik}, R.~P., {Szklen{\'a}r}, T., {Berlind}, P.,
  {Calkins}, M.~L., {L{\'a}z{\'a}r}, J., {Papp}, I., \& {S{\'a}ri}, P. 2011,
  \apj, 733, 116

\bibitem[{{Bundy} \& {Marcy}(2000)}]{Bundy_Marcy_00}
{Bundy}, K.~A. \& {Marcy}, G.~W. 2000, \pasp, 112, 1421

\bibitem[{{Burke} {et~al.}(2006){Burke}, {Gaudi}, {DePoy}, \&
  {Pogge}}]{Burke_et_al_06}
{Burke}, C.~J., {Gaudi}, B.~S., {DePoy}, D.~L., \& {Pogge}, R.~W. 2006, \aj,
  132, 210

\bibitem[{{Burke} {et~al.}(2007){Burke}, {McCullough}, {Valenti},
  {Johns-Krull}, {Janes}, {Heasley}, {Summers}, {Stys}, {Bissinger}, {Fleenor},
  {Foote}, {Garc{\'{\i}}a-Melendo}, {Gary}, {Howell}, {Mallia}, {Masi},
  {Taylor}, \& {Vanmunster}}]{Burke_et_al_07}
{Burke}, C.~J., {McCullough}, P.~R., {Valenti}, J.~A., {Johns-Krull}, C.~M.,
  {Janes}, K.~A., {Heasley}, J.~N., {Summers}, F.~J., {Stys}, J.~E.,
  {Bissinger}, R., {Fleenor}, M.~L., {Foote}, C.~N., {Garc{\'{\i}}a-Melendo},
  E., {Gary}, B.~L., {Howell}, P.~J., {Mallia}, F., {Masi}, G., {Taylor}, B.,
  \& {Vanmunster}, T. 2007, \apj, 671, 2115

\bibitem[{{Carone} \& {P{\"a}tzold}(2007)}]{Carone_Patzold_07}
{Carone}, L. \& {P{\"a}tzold}, M. 2007, \planss, 55, 643

\bibitem[{{Chan} {et~al.}(2011){Chan}, {Ingemyr}, {Winn}, {Holman},
  {Sanchis-Ojeda}, {Esquerdo}, \& {Everett}}]{Chan_et_al_11}
{Chan}, T., {Ingemyr}, M., {Winn}, J.~N., {Holman}, M.~J., {Sanchis-Ojeda}, R.,
  {Esquerdo}, G., \& {Everett}, M. 2011, \aj, 141, 179

\bibitem[{{Christian} {et~al.}(2009){Christian}, {Gibson}, {Simpson}, {Street},
  {Skillen}, {Pollacco}, {Collier Cameron}, {Joshi}, {Keenan}, {Stempels},
  {Haswell}, {Horne}, {Anderson}, {Bentley}, {Bouchy}, {Clarkson}, {Enoch},
  {Hebb}, {H{\'e}brard}, {Hellier}, {Irwin}, {Kane}, {Lister}, {Loeillet},
  {Maxted}, {Mayor}, {McDonald}, {Moutou}, {Norton}, {Parley}, {Pont},
  {Queloz}, {Ryans}, {Smalley}, {Smith}, {Todd}, {Udry}, {West}, {Wheatley}, \&
  {Wilson}}]{Christian_et_al_09}
{Christian}, D.~J., {Gibson}, N.~P., {Simpson}, E.~K., {Street}, R.~A.,
  {Skillen}, I., {Pollacco}, D., {Collier Cameron}, A., {Joshi}, Y.~C.,
  {Keenan}, F.~P., {Stempels}, H.~C., {Haswell}, C.~A., {Horne}, K.,
  {Anderson}, D.~R., {Bentley}, S., {Bouchy}, F., {Clarkson}, W.~I., {Enoch},
  B., {Hebb}, L., {H{\'e}brard}, G., {Hellier}, C., {Irwin}, J., {Kane}, S.~R.,
  {Lister}, T.~A., {Loeillet}, B., {Maxted}, P., {Mayor}, M., {McDonald}, I.,
  {Moutou}, C., {Norton}, A.~J., {Parley}, N., {Pont}, F., {Queloz}, D.,
  {Ryans}, R., {Smalley}, B., {Smith}, A.~M.~S., {Todd}, I., {Udry}, S.,
  {West}, R.~G., {Wheatley}, P.~J., \& {Wilson}, D.~M. 2009, \mnras, 392, 1585

\bibitem[{{Collier Cameron} {et~al.}(2006){Collier Cameron}, {Pollacco},
  {Street}, {Lister}, {West}, {Wilson}, {Pont}, {Christian}, {Clarkson},
  {Enoch}, {Evans}, {Fitzsimmons}, {Haswell}, {Hellier}, {Hodgkin}, {Horne},
  {Irwin}, {Kane}, {Keenan}, {Norton}, {Parley}, {Osborne}, {Ryans}, {Skillen},
  \& {Wheatley}}]{Collier_et_al_06}
{Collier Cameron}, A., {Pollacco}, D., {Street}, R.~A., {Lister}, T.~A.,
  {West}, R.~G., {Wilson}, D.~M., {Pont}, F., {Christian}, D.~J., {Clarkson},
  W.~I., {Enoch}, B., {Evans}, A., {Fitzsimmons}, A., {Haswell}, C.~A.,
  {Hellier}, C., {Hodgkin}, S.~T., {Horne}, K., {Irwin}, J., {Kane}, S.~R.,
  {Keenan}, F.~P., {Norton}, A.~J., {Parley}, N.~R., {Osborne}, J., {Ryans},
  R., {Skillen}, I., \& {Wheatley}, P.~J. 2006, \mnras, 373, 799

\bibitem[{{Counselman}(1973)}]{Counselman_73}
{Counselman}, III, C.~C. 1973, \apj, 180, 307

\bibitem[{{Daemgen} {et~al.}(2009){Daemgen}, {Hormuth}, {Brandner}, {Bergfors},
  {Janson}, {Hippler}, \& {Henning}}]{Daemgen_et_al_09}
{Daemgen}, S., {Hormuth}, F., {Brandner}, W., {Bergfors}, C., {Janson}, M.,
  {Hippler}, S., \& {Henning}, T. 2009, \aap, 498, 567

\bibitem[{{Demarque} {et~al.}(2008){Demarque}, {Guenther}, {Li}, {Mazumdar}, \&
  {Straka}}]{Demarque_et_al_08}
{Demarque}, P., {Guenther}, D.~B., {Li}, L.~H., {Mazumdar}, A., \& {Straka},
  C.~W. 2008, \apss, 316, 31

\bibitem[{{Denissenkov}(2010)}]{Denissenkov_10}
{Denissenkov}, P.~A. 2010, \apj, 719, 28

\bibitem[{{Enoch} {et~al.}(2011{\natexlab{a}}){Enoch}, {Anderson}, {Barros},
  {Brown}, {Collier Cameron}, {Faedi}, {Gillon}, {H{\'e}brard}, {Lister},
  {Queloz}, {Santerne}, {Smalley}, {Street}, {Triaud}, {West}, {Bouchy},
  {Bento}, {Butters}, {Fossati}, {Haswell}, {Hellier}, {Holmes}, {Jehen},
  {Lendl}, {Maxted}, {McCormac}, {Miller}, {Moulds}, {Moutou}, {Norton},
  {Parley}, {Pepe}, {Pollacco}, {Segransan}, {Simpson}, {Skillen}, {Smith},
  {Udry}, \& {Wheatley}}]{Enoch_et_al_11b}
{Enoch}, B., {Anderson}, D.~R., {Barros}, S.~C.~C., {Brown}, D.~J.~A., {Collier
  Cameron}, A., {Faedi}, F., {Gillon}, M., {H{\'e}brard}, G., {Lister}, T.~A.,
  {Queloz}, D., {Santerne}, A., {Smalley}, B., {Street}, R.~A., {Triaud},
  A.~H.~M.~J., {West}, R.~G., {Bouchy}, F., {Bento}, J., {Butters}, O.,
  {Fossati}, L., {Haswell}, C.~A., {Hellier}, C., {Holmes}, S., {Jehen}, E.,
  {Lendl}, M., {Maxted}, P.~F.~L., {McCormac}, J., {Miller}, G.~R.~M.,
  {Moulds}, V., {Moutou}, C., {Norton}, A.~J., {Parley}, N., {Pepe}, F.,
  {Pollacco}, D., {Segransan}, D., {Simpson}, E., {Skillen}, I., {Smith},
  A.~M.~S., {Udry}, S., \& {Wheatley}, P.~J. 2011{\natexlab{a}}, \aj, 142, 86

\bibitem[{{Enoch} {et~al.}(2011{\natexlab{b}}){Enoch}, {Cameron}, {Anderson},
  {Lister}, {Hellier}, {Maxted}, {Queloz}, {Smalley}, {Triaud}, {West},
  {Brown}, {Gillon}, {Hebb}, {Lendl}, {Parley}, {Pepe}, {Pollacco},
  {Segransan}, {Simpson}, {Street}, \& {Udry}}]{Enoch_et_al_11a}
{Enoch}, B., {Cameron}, A.~C., {Anderson}, D.~R., {Lister}, T.~A., {Hellier},
  C., {Maxted}, P.~F.~L., {Queloz}, D., {Smalley}, B., {Triaud}, A.~H.~M.~J.,
  {West}, R.~G., {Brown}, D.~J.~A., {Gillon}, M., {Hebb}, L., {Lendl}, M.,
  {Parley}, N., {Pepe}, F., {Pollacco}, D., {Segransan}, D., {Simpson}, E.,
  {Street}, R.~A., \& {Udry}, S. 2011{\natexlab{b}}, \mnras, 410, 1631

\bibitem[{{Faedi} {et~al.}(2011){Faedi}, {Barros}, {Anderson}, {Brown},
  {Collier Cameron}, {Pollacco}, {Boisse}, {H{\'e}brard}, {Lendl}, {Lister},
  {Smalley}, {Street}, {Triaud}, {Bento}, {Bouchy}, {Butters}, {Enoch},
  {Haswell}, {Hellier}, {Keenan}, {Miller}, {Moulds}, {Moutou}, {Norton},
  {Queloz}, {Santerne}, {Simpson}, {Skillen}, {Smith}, {Udry}, {Watson},
  {West}, \& {Wheatley}}]{Faedi_et_al_11}
{Faedi}, F., {Barros}, S.~C.~C., {Anderson}, D.~R., {Brown}, D.~J.~A., {Collier
  Cameron}, A., {Pollacco}, D., {Boisse}, I., {H{\'e}brard}, G., {Lendl}, M.,
  {Lister}, T.~A., {Smalley}, B., {Street}, R.~A., {Triaud}, A.~H.~M.~J.,
  {Bento}, J., {Bouchy}, F., {Butters}, O.~W., {Enoch}, B., {Haswell}, C.~A.,
  {Hellier}, C., {Keenan}, F.~P., {Miller}, G.~R.~M., {Moulds}, V., {Moutou},
  C., {Norton}, A.~J., {Queloz}, D., {Santerne}, A., {Simpson}, E.~K.,
  {Skillen}, I., {Smith}, A.~M.~S., {Udry}, S., {Watson}, C.~A., {West}, R.~G.,
  \& {Wheatley}, P.~J. 2011, \aap, 531, A40+

\bibitem[{{Fischer} {et~al.}(2008){Fischer}, {Marcy}, {Butler}, {Vogt},
  {Laughlin}, {Henry}, {Abouav}, {Peek}, {Wright}, {Johnson}, {McCarthy}, \&
  {Isaacson}}]{Fischer_et_al_08}
{Fischer}, D.~A., {Marcy}, G.~W., {Butler}, R.~P., {Vogt}, S.~S., {Laughlin},
  G., {Henry}, G.~W., {Abouav}, D., {Peek}, K.~M.~G., {Wright}, J.~T.,
  {Johnson}, J.~A., {McCarthy}, C., \& {Isaacson}, H. 2008, \apj, 675, 790

\bibitem[{{Gillon} {et~al.}(2006){Gillon}, {Pont}, {Moutou}, {Bouchy},
  {Courbin}, {Sohy}, \& {Magain}}]{Gillon_et_al_06}
{Gillon}, M., {Pont}, F., {Moutou}, C., {Bouchy}, F., {Courbin}, F., {Sohy},
  S., \& {Magain}, P. 2006, \aap, 459, 249

\bibitem[{{Goldreich}(1963)}]{Goldreich_63}
{Goldreich}, P. 1963, \mnras, 126, 257

\bibitem[{{Goldreich} \& {Keeley}(1977)}]{Goldreich_Keeley_77}
{Goldreich}, P. \& {Keeley}, D.~A. 1977, \apj, 211, 934

\bibitem[{{Goldreich} \& {Nicholson}(1977)}]{Goldreich_Nicholson_77}
{Goldreich}, P. \& {Nicholson}, P.~D. 1977, Icarus, 30, 301

\bibitem[{{Goodman} \& {Oh}(1997)}]{Goodman_Oh_97}
{Goodman}, J. \& {Oh}, S.~P. 1997, \apj, 486, 403

\bibitem[{{Greenberg}(1974)}]{Greenberg_74}
{Greenberg}, R. 1974, Icarus, 23, 51

\bibitem[{{Haisch} {et~al.}(2001){Haisch}, {Lada}, \&
  {Lada}}]{Haisch_Lada_Lada_01}
{Haisch}, Jr., K.~E., {Lada}, E.~A., \& {Lada}, C.~J. 2001, \apjl, 553, L153

\bibitem[{{Hansen}(2010)}]{Hansen_10}
{Hansen}, B.~M.~S. 2010, \apj, 723, 285

\bibitem[{{Hartman} {et~al.}(2011{\natexlab{a}}){Hartman}, {Bakos}, {Kipping},
  {Torres}, {Kov{\'a}cs}, {Noyes}, {Latham}, {Howard}, {Fischer}, {Johnson},
  {Marcy}, {Isaacson}, {Quinn}, {Buchhave}, {B{\'e}ky}, {Sasselov}, {Stefanik},
  {Esquerdo}, {Everett}, {Perumpilly}, {L{\'a}z{\'a}r}, {Papp}, \&
  {S{\'a}ri}}]{Hartman_et_al_11b}
{Hartman}, J.~D., {Bakos}, G.~{\'A}., {Kipping}, D.~M., {Torres}, G.,
  {Kov{\'a}cs}, G., {Noyes}, R.~W., {Latham}, D.~W., {Howard}, A.~W.,
  {Fischer}, D.~A., {Johnson}, J.~A., {Marcy}, G.~W., {Isaacson}, H., {Quinn},
  S.~N., {Buchhave}, L.~A., {B{\'e}ky}, B., {Sasselov}, D.~D., {Stefanik},
  R.~P., {Esquerdo}, G.~A., {Everett}, M., {Perumpilly}, G., {L{\'a}z{\'a}r},
  J., {Papp}, I., \& {S{\'a}ri}, P. 2011{\natexlab{a}}, \apj, 728, 138

\bibitem[{{Hartman} {et~al.}(2011{\natexlab{b}}){Hartman}, {Bakos}, {Sato},
  {Torres}, {Noyes}, {Latham}, {Kov{\'a}cs}, {Fischer}, {Howard}, {Johnson},
  {Marcy}, {Buchhave}, {F{\"u}resz}, {Perumpilly}, {B{\'e}ky}, {Stefanik},
  {Sasselov}, {Esquerdo}, {Everett}, {Csubry}, {L{\'a}z{\'a}r}, {Papp}, \&
  {S{\'a}ri}}]{Hartman_et_al_11a}
{Hartman}, J.~D., {Bakos}, G.~{\'A}., {Sato}, B., {Torres}, G., {Noyes}, R.~W.,
  {Latham}, D.~W., {Kov{\'a}cs}, G., {Fischer}, D.~A., {Howard}, A.~W.,
  {Johnson}, J.~A., {Marcy}, G.~W., {Buchhave}, L.~A., {F{\"u}resz}, G.,
  {Perumpilly}, G., {B{\'e}ky}, B., {Stefanik}, R.~P., {Sasselov}, D.~D.,
  {Esquerdo}, G.~A., {Everett}, M., {Csubry}, Z., {L{\'a}z{\'a}r}, J., {Papp},
  I., \& {S{\'a}ri}, P. 2011{\natexlab{b}}, \apj, 726, 52

\bibitem[{{Hartman} {et~al.}(2009{\natexlab{a}}){Hartman}, {Bakos}, {Torres},
  {Kov{\'a}cs}, {Noyes}, {P{\'a}l}, {Latham}, {Sip{\H o}cz}, {Fischer},
  {Johnson}, {Marcy}, {Butler}, {Howard}, {Esquerdo}, {Sasselov}, {Kov{\'a}cs},
  {Stefanik}, {Fernandez}, {L{\'a}z{\'a}r}, {Papp}, \&
  {S{\'a}ri}}]{Hartman_et_al_09b}
{Hartman}, J.~D., {Bakos}, G.~{\'A}., {Torres}, G., {Kov{\'a}cs}, G., {Noyes},
  R.~W., {P{\'a}l}, A., {Latham}, D.~W., {Sip{\H o}cz}, B., {Fischer}, D.~A.,
  {Johnson}, J.~A., {Marcy}, G.~W., {Butler}, R.~P., {Howard}, A.~W.,
  {Esquerdo}, G.~A., {Sasselov}, D.~D., {Kov{\'a}cs}, G., {Stefanik}, R.~P.,
  {Fernandez}, J.~M., {L{\'a}z{\'a}r}, J., {Papp}, I., \& {S{\'a}ri}, P.
  2009{\natexlab{a}}, \apj, 706, 785

\bibitem[{{Hartman} {et~al.}(2011{\natexlab{c}}){Hartman}, {Bakos}, {Torres},
  {Latham}, {Kov{\'a}cs}, {B{\'e}ky}, {Quinn}, {Mazeh}, {Shporer}, {Marcy},
  {Howard}, {Fischer}, {Johnson}, {Esquerdo}, {Noyes}, {Sasselov}, {Stefanik},
  {Fernandez}, {Szklen{\'a}r}, {L{\'a}z{\'a}r}, {Papp}, \&
  {S{\'a}ri}}]{Hartman_et_al_11c}
{Hartman}, J.~D., {Bakos}, G.~{\'A}., {Torres}, G., {Latham}, D.~W.,
  {Kov{\'a}cs}, G., {B{\'e}ky}, B., {Quinn}, S.~N., {Mazeh}, T., {Shporer}, A.,
  {Marcy}, G.~W., {Howard}, A.~W., {Fischer}, D.~A., {Johnson}, J.~A.,
  {Esquerdo}, G.~A., {Noyes}, R.~W., {Sasselov}, D.~D., {Stefanik}, R.~P.,
  {Fernandez}, J.~M., {Szklen{\'a}r}, T., {L{\'a}z{\'a}r}, J., {Papp}, I., \&
  {S{\'a}ri}, P. 2011{\natexlab{c}}, \apj, 742, 59

\bibitem[{{Hartman} {et~al.}(2009{\natexlab{b}}){Hartman}, {Gaudi}, {Holman},
  {McLeod}, {Stanek}, {Barranco}, {Pinsonneault}, {Meibom}, \&
  {Kalirai}}]{Hartman_et_al_09a}
{Hartman}, J.~D., {Gaudi}, B.~S., {Holman}, M.~J., {McLeod}, B.~A., {Stanek},
  K.~Z., {Barranco}, J.~A., {Pinsonneault}, M.~H., {Meibom}, S., \& {Kalirai},
  J.~S. 2009{\natexlab{b}}, \apj, 695, 336

\bibitem[{{Hellier} {et~al.}(2009){Hellier}, {Anderson}, {Collier Cameron},
  {Gillon}, {Hebb}, {Maxted}, {Queloz}, {Smalley}, {Triaud}, {West}, {Wilson},
  {Bentley}, {Enoch}, {Horne}, {Irwin}, {Lister}, {Mayor}, {Parley}, {Pepe},
  {Pollacco}, {Segransan}, {Udry}, \& {Wheatley}}]{Hellier_et_al_09}
{Hellier}, C., {Anderson}, D.~R., {Collier Cameron}, A., {Gillon}, M., {Hebb},
  L., {Maxted}, P.~F.~L., {Queloz}, D., {Smalley}, B., {Triaud}, A.~H.~M.~J.,
  {West}, R.~G., {Wilson}, D.~M., {Bentley}, S.~J., {Enoch}, B., {Horne}, K.,
  {Irwin}, J., {Lister}, T.~A., {Mayor}, M., {Parley}, N., {Pepe}, F.,
  {Pollacco}, D.~L., {Segransan}, D., {Udry}, S., \& {Wheatley}, P.~J. 2009,
  \nat, 460, 1098

\bibitem[{{Hellier} {et~al.}(2011{\natexlab{a}}){Hellier}, {Anderson}, {Collier
  Cameron}, {Gillon}, {Jehin}, {Lendl}, {Maxted}, {Pepe}, {Pollacco}, {Queloz},
  {S{\'e}gransan}, {Smalley}, {Smith}, {Southworth}, {Triaud}, {Udry}, \&
  {West}}]{Hellier_et_al_11b}
{Hellier}, C., {Anderson}, D.~R., {Collier Cameron}, A., {Gillon}, M., {Jehin},
  E., {Lendl}, M., {Maxted}, P.~F.~L., {Pepe}, F., {Pollacco}, D., {Queloz},
  D., {S{\'e}gransan}, D., {Smalley}, B., {Smith}, A.~M.~S., {Southworth}, J.,
  {Triaud}, A.~H.~M.~J., {Udry}, S., \& {West}, R.~G. 2011{\natexlab{a}}, \aap,
  535, L7

\bibitem[{{Hellier} {et~al.}(2011{\natexlab{b}}){Hellier}, {Anderson},
  {Collier-Cameron}, {Miller}, {Queloz}, {Smalley}, {Southworth}, \&
  {Triaud}}]{Hellier_et_al_11a}
{Hellier}, C., {Anderson}, D.~R., {Collier-Cameron}, A., {Miller}, G.~R.~M.,
  {Queloz}, D., {Smalley}, B., {Southworth}, J., \& {Triaud}, A.~H.~M.~J.
  2011{\natexlab{b}}, \apjl, 730, L31+

\bibitem[{{Ibgui} \& {Burrows}(2009)}]{Ibgui_Burrows_09}
{Ibgui}, L. \& {Burrows}, A. 2009, \apj, 700, 1921

\bibitem[{{Irwin} \& {Bouvier}(2009)}]{Irwin_Bouvier_09}
{Irwin}, J. \& {Bouvier}, J. 2009, in IAU Symposium, Vol. 258, IAU Symposium,
  ed. {E.~E.~Mamajek, D.~R.~Soderblom, \& R.~F.~G.~Wyse}, 363--374

\bibitem[{{Irwin} {et~al.}(2007){Irwin}, {Hodgkin}, {Aigrain}, {Hebb},
  {Bouvier}, {Clarke}, {Moraux}, \& {Bramich}}]{Irwin_et_al_07}
{Irwin}, J., {Hodgkin}, S., {Aigrain}, S., {Hebb}, L., {Bouvier}, J., {Clarke},
  C., {Moraux}, E., \& {Bramich}, D.~M. 2007, \mnras, 377, 741

\bibitem[{{Ivanov} \& {Papaloizou}(2007)}]{Ivanov_Papaloizou_07}
{Ivanov}, P.~B. \& {Papaloizou}, J.~C.~B. 2007, \mnras, 376, 682

\bibitem[{{Jackson} {et~al.}(2008{\natexlab{a}}){Jackson}, {Greenberg}, \&
  {Barnes}}]{Jackson_et_al_08a}
{Jackson}, B., {Greenberg}, R., \& {Barnes}, R. 2008{\natexlab{a}}, \apj, 678,
  1396

\bibitem[{{Jackson} {et~al.}(2008{\natexlab{b}}){Jackson}, {Greenberg}, \&
  {Barnes}}]{Jackson_et_al_08b}
---. 2008{\natexlab{b}}, \apj, 681, 1631

\bibitem[{{Johnson} {et~al.}(2009){Johnson}, {Winn}, {Albrecht}, {Howard},
  {Marcy}, \& {Gazak}}]{Johnson_et_al_09}
{Johnson}, J.~A., {Winn}, J.~N., {Albrecht}, S., {Howard}, A.~W., {Marcy},
  G.~W., \& {Gazak}, J.~Z. 2009, \pasp, 121, 1104

\bibitem[{{Johnson} {et~al.}(2011){Johnson}, {Winn}, {Bakos}, {Hartman},
  {Morton}, {Torres}, {Kov{\'a}cs}, {Latham}, {Noyes}, {Sato}, {Esquerdo},
  {Fischer}, {Marcy}, {Howard}, {Buchhave}, {F{\H u}r{\'e}sz}, {Quinn},
  {B{\'e}ky}, {Sasselov}, {Stefanik}, {L{\'a}z{\'a}r}, {Papp}, \&
  {S{\'a}ri}}]{Johnson_et_al_11}
{Johnson}, J.~A., {Winn}, J.~N., {Bakos}, G.~{\'A}., {Hartman}, J.~D.,
  {Morton}, T.~D., {Torres}, G., {Kov{\'a}cs}, G., {Latham}, D.~W., {Noyes},
  R.~W., {Sato}, B., {Esquerdo}, G.~A., {Fischer}, D.~A., {Marcy}, G.~W.,
  {Howard}, A.~W., {Buchhave}, L.~A., {F{\H u}r{\'e}sz}, G., {Quinn}, S.~N.,
  {B{\'e}ky}, B., {Sasselov}, D.~D., {Stefanik}, R.~P., {L{\'a}z{\'a}r}, J.,
  {Papp}, I., \& {S{\'a}ri}, P. 2011, \apj, 735, 24

\bibitem[{{Johnson} {et~al.}(2008){Johnson}, {Winn}, {Narita}, {Enya},
  {Williams}, {Marcy}, {Sato}, {Ohta}, {Taruya}, {Suto}, {Turner}, {Bakos},
  {Butler}, {Vogt}, {Aoki}, {Tamura}, {Yamada}, {Yoshii}, \&
  {Hidas}}]{Johnson_et_al_08}
{Johnson}, J.~A., {Winn}, J.~N., {Narita}, N., {Enya}, K., {Williams},
  P.~K.~G., {Marcy}, G.~W., {Sato}, B., {Ohta}, Y., {Taruya}, A., {Suto}, Y.,
  {Turner}, E.~L., {Bakos}, G., {Butler}, R.~P., {Vogt}, S.~S., {Aoki}, W.,
  {Tamura}, M., {Yamada}, T., {Yoshii}, Y., \& {Hidas}, M. 2008, \apj, 686, 649

\bibitem[{{Kaula}(1968)}]{Kaula_68}
{Kaula}, W.~M. 1968, {An introduction to planetary physics - The terrestrial
  planets}, ed. {Kaula, W.~M.}

\bibitem[{{Kawaler}(1988)}]{Kawaler_88}
{Kawaler}, S.~D. 1988, \apj, 333, 236

\bibitem[{{Kipping} {et~al.}(2010){Kipping}, {Bakos}, {Hartman}, {Torres},
  {Shporer}, {Latham}, {Kov{\'a}cs}, {Noyes}, {Howard}, {Fischer}, {Johnson},
  {Marcy}, {B{\'e}ky}, {Perumpilly}, {Esquerdo}, {Sasselov}, {Stefanik},
  {L{\'a}z{\'a}r}, {Papp}, \& {S{\'a}ri}}]{Kipping_et_al_10}
{Kipping}, D.~M., {Bakos}, G.~{\'A}., {Hartman}, J., {Torres}, G., {Shporer},
  A., {Latham}, D.~W., {Kov{\'a}cs}, G., {Noyes}, R.~W., {Howard}, A.~W.,
  {Fischer}, D.~A., {Johnson}, J.~A., {Marcy}, G.~W., {B{\'e}ky}, B.,
  {Perumpilly}, G., {Esquerdo}, G.~A., {Sasselov}, D.~D., {Stefanik}, R.~P.,
  {L{\'a}z{\'a}r}, J., {Papp}, I., \& {S{\'a}ri}, P. 2010, \apj, 725, 2017

\bibitem[{{Lai}(2012)}]{Lai_12}
{Lai}, D. 2012, \mnras, 2789

\bibitem[{{Levrard} {et~al.}(2009){Levrard}, {Winisdoerffer}, \&
  {Chabrier}}]{Levrard_et_al_09}
{Levrard}, B., {Winisdoerffer}, C., \& {Chabrier}, G. 2009, \apjl, 692, L9

\bibitem[{{Lin} {et~al.}(1996){Lin}, {Bodenheimer}, \&
  {Richardson}}]{Lin_et_al_96}
{Lin}, D.~N.~C., {Bodenheimer}, P., \& {Richardson}, D.~C. 1996, \nat, 380, 606

\bibitem[{{Lister} {et~al.}(2009){Lister}, {Anderson}, {Gillon}, {Hebb},
  {Smalley}, {Triaud}, {Collier Cameron}, {Wilson}, {West}, {Bentley},
  {Christian}, {Enoch}, {Haswell}, {Hellier}, {Horne}, {Irwin}, {Joshi},
  {Kane}, {Mayor}, {Maxted}, {Norton}, {Parley}, {Pepe}, {Pollacco}, {Queloz},
  {Ryans}, {Segransan}, {Skillen}, {Street}, {Todd}, {Udry}, \&
  {Wheatley}}]{Lister_et_al_09}
{Lister}, T.~A., {Anderson}, D.~R., {Gillon}, M., {Hebb}, L., {Smalley}, B.~S.,
  {Triaud}, A.~H.~M.~J., {Collier Cameron}, A., {Wilson}, D.~M., {West}, R.~G.,
  {Bentley}, S.~J., {Christian}, D.~J., {Enoch}, R., {Haswell}, C.~A.,
  {Hellier}, C., {Horne}, K., {Irwin}, J., {Joshi}, Y.~C., {Kane}, S.~R.,
  {Mayor}, M., {Maxted}, P.~F.~L., {Norton}, A.~J., {Parley}, N., {Pepe}, F.,
  {Pollacco}, D., {Queloz}, D., {Ryans}, R., {Segransan}, D., {Skillen}, I.,
  {Street}, R.~A., {Todd}, I., {Udry}, S., \& {Wheatley}, P.~J. 2009, \apj,
  703, 752

\bibitem[{{Liu} {et~al.}(2008){Liu}, {Burrows}, \& {Ibgui}}]{Liu_et_al_08}
{Liu}, X., {Burrows}, A., \& {Ibgui}, L. 2008, \apj, 687, 1191

\bibitem[{{MacGregor}(1991)}]{MacGregor_91}
{MacGregor}, K.~B. 1991, in NATO ASIC Proc. 340: Angular Momentum Evolution of
  Young Stars, 315--+

\bibitem[{{Maxted} {et~al.}(2011){Maxted}, {Anderson}, {Collier Cameron},
  {Hellier}, {Queloz}, {Smalley}, {Street}, {Triaud}, {West}, {Gillon},
  {Lister}, {Pepe}, {Pollacco}, {S{\'e}gransan}, {Smith}, \&
  {Udry}}]{Maxted_et_al_11}
{Maxted}, P.~F.~L., {Anderson}, D.~R., {Collier Cameron}, A., {Hellier}, C.,
  {Queloz}, D., {Smalley}, B., {Street}, R.~A., {Triaud}, A.~H.~M.~J., {West},
  R.~G., {Gillon}, M., {Lister}, T.~A., {Pepe}, F., {Pollacco}, D.,
  {S{\'e}gransan}, D., {Smith}, A.~M.~S., \& {Udry}, S. 2011, \pasp, 123, 547

\bibitem[{{Maxted} {et~al.}(2010){Maxted}, {Anderson}, {Gillon}, {Hellier},
  {Queloz}, {Smalley}, {Triaud}, {West}, {Wilson}, {Bentley}, {Cegla}, {Collier
  Cameron}, {Enoch}, {Hebb}, {Horne}, {Irwin}, {Lister}, {Mayor}, {Parley},
  {Pepe}, {Pollacco}, {Segransan}, {Udry}, \& {Wheatley}}]{Maxted_et_al_10}
{Maxted}, P.~F.~L., {Anderson}, D.~R., {Gillon}, M., {Hellier}, C., {Queloz},
  D., {Smalley}, B., {Triaud}, A.~H.~M.~J., {West}, R.~G., {Wilson}, D.~M.,
  {Bentley}, S.~J., {Cegla}, H., {Collier Cameron}, A., {Enoch}, B., {Hebb},
  L., {Horne}, K., {Irwin}, J., {Lister}, T.~A., {Mayor}, M., {Parley}, N.,
  {Pepe}, F., {Pollacco}, D., {Segransan}, D., {Udry}, S., \& {Wheatley}, P.~J.
  2010, \aj, 140, 2007

\bibitem[{{Meibom} \& {Mathieu}(2005)}]{Meibom_Mathieu_05}
{Meibom}, S. \& {Mathieu}, R.~D. 2005, \apj, 620, 970

\bibitem[{{Miller} {et~al.}(2009){Miller}, {Fortney}, \&
  {Jackson}}]{Miller_et_al_09}
{Miller}, N., {Fortney}, J.~J., \& {Jackson}, B. 2009, \apj, 702, 1413

\bibitem[{{Narita} {et~al.}(2007){Narita}, {Enya}, {Sato}, {Ohta}, {Winn},
  {Suto}, {Taruya}, {Turner}, {Aoki}, {Yoshii}, {Yamada}, \&
  {Tamura}}]{Narita_et_al_07}
{Narita}, N., {Enya}, K., {Sato}, B., {Ohta}, Y., {Winn}, J.~N., {Suto}, Y.,
  {Taruya}, A., {Turner}, E.~L., {Aoki}, W., {Yoshii}, M., {Yamada}, T., \&
  {Tamura}, Y. 2007, \pasj, 59, 763

\bibitem[{{Narita} {et~al.}(2010{\natexlab{a}}){Narita}, {Hirano},
  {Sanchis-Ojeda}, {Winn}, {Holman}, {Sato}, {Aoki}, \&
  {Tamura}}]{Narita_et_al_10b}
{Narita}, N., {Hirano}, T., {Sanchis-Ojeda}, R., {Winn}, J.~N., {Holman},
  M.~J., {Sato}, B., {Aoki}, W., \& {Tamura}, M. 2010{\natexlab{a}}, \pasj, 62,
  L61

\bibitem[{{Narita} {et~al.}(2010{\natexlab{b}}){Narita}, {Sato}, {Hirano},
  {Winn}, {Aoki}, \& {Tamura}}]{Narita_et_al_10a}
{Narita}, N., {Sato}, B., {Hirano}, T., {Winn}, J.~N., {Aoki}, W., \& {Tamura},
  M. 2010{\natexlab{b}}, \pasj, 62, 653

\bibitem[{{Narita} {et~al.}(2008){Narita}, {Sato}, {Ohshima}, \&
  {Winn}}]{Narita_et_al_08}
{Narita}, N., {Sato}, B., {Ohshima}, O., \& {Winn}, J.~N. 2008, \pasj, 60, L1

\bibitem[{{Ogilvie}(2009)}]{Ogilvie_09}
{Ogilvie}, G.~I. 2009, \mnras, 396, 794

\bibitem[{{Ogilvie} \& {Lin}(2004)}]{Ogilvie_Lin_04}
{Ogilvie}, G.~I. \& {Lin}, D.~N.~C. 2004, \apj, 610, 477

\bibitem[{{Ogilvie} \& {Lin}(2007)}]{Ogilvie_Lin_07}
---. 2007, \apj, 661, 1180

\bibitem[{{Papaloizou} {et~al.}(1997){Papaloizou}, {Alberts}, {Pringle}, \&
  {Savonije}}]{Savonije_Papaloizou_97c}
{Papaloizou}, J.~C.~B., {Alberts}, F., {Pringle}, J.~E., \& {Savonije}, G.~J.
  1997, \mnras, 284, 821

\bibitem[{{Papaloizou} \& {Ivanov}(2005)}]{Papaloizou_Ivanov_05}
{Papaloizou}, J.~C.~B. \& {Ivanov}, P.~B. 2005, \mnras, 364, L66

\bibitem[{{Papaloizou} \& {Savonije}(1997)}]{Savonije_Papaloizou_97a}
{Papaloizou}, J.~C.~B. \& {Savonije}, G.~J. 1997, \mnras, 291, 651

\bibitem[{{Penev} {et~al.}(2009{\natexlab{a}}){Penev}, {Barranco}, \&
  {Sasselov}}]{Penev_Barranco_Sasselov_09}
{Penev}, K., {Barranco}, J., \& {Sasselov}, D. 2009{\natexlab{a}}, \apj, 705,
  285

\bibitem[{{Penev} {et~al.}(2011){Penev}, {Barranco}, \&
  {Sasselov}}]{Penev_Barranco_Sasselov_11}
---. 2011, \apj, 734, 118

\bibitem[{{Penev} \& {Sasselov}(2011)}]{Penev_Sasselov_11}
{Penev}, K. \& {Sasselov}, D. 2011, \apj, 731, 67

\bibitem[{{Penev} {et~al.}(2007){Penev}, {Sasselov}, {Robinson}, \&
  {Demarque}}]{Penev_Sasselov_Robinson_Demarque_07}
{Penev}, K., {Sasselov}, D., {Robinson}, F., \& {Demarque}, P. 2007, \apj, 655,
  1166

\bibitem[{{Penev} {et~al.}(2009{\natexlab{b}}){Penev}, {Sasselov}, {Robinson},
  \& {Demarque}}]{Penev_Sasselov_Robinson_Demarque_09}
---. 2009{\natexlab{b}}, \apj, 704, 930

\bibitem[{{Pont} {et~al.}(2007){Pont}, {Moutou}, {Gillon}, {Udalski}, {Bouchy},
  {Fernandes}, {Gieren}, {Mayor}, {Mazeh}, {Minniti}, {Melo}, {Naef},
  {Pietrzynski}, {Queloz}, {Ruiz}, {Santos}, \& {Udry}}]{Pont_et_al_07}
{Pont}, F., {Moutou}, C., {Gillon}, M., {Udalski}, A., {Bouchy}, F.,
  {Fernandes}, J.~M., {Gieren}, W., {Mayor}, M., {Mazeh}, T., {Minniti}, D.,
  {Melo}, C., {Naef}, D., {Pietrzynski}, G., {Queloz}, D., {Ruiz}, M.~T.,
  {Santos}, N.~C., \& {Udry}, S. 2007, \aap, 465, 1069

\bibitem[{{Queloz} {et~al.}(2000){Queloz}, {Eggenberger}, {Mayor}, {Perrier},
  {Beuzit}, {Naef}, {Sivan}, \& {Udry}}]{Queloz_et_al_00}
{Queloz}, D., {Eggenberger}, A., {Mayor}, M., {Perrier}, C., {Beuzit}, J.~L.,
  {Naef}, D., {Sivan}, J.~P., \& {Udry}, S. 2000, \aap, 359, L13

\bibitem[{{Quinn} {et~al.}(2012){Quinn}, {Bakos}, {Hartman}, {Torres},
  {Kov{\'a}cs}, {Latham}, {Noyes}, {Fischer}, {Johnson}, {Marcy}, {Howard},
  {Szentgyorgyi}, {F{\H u}r{\'e}sz}, {Buchhave}, {B{\'e}ky}, {Sasselov},
  {Stefanik}, {Perumpilly}, {Everett}, {L{\'a}z{\'a}r}, {Papp}, \&
  {S{\'a}ri}}]{Quinn_et_al_12}
{Quinn}, S.~N., {Bakos}, G.~{\'A}., {Hartman}, J., {Torres}, G., {Kov{\'a}cs},
  G., {Latham}, D.~W., {Noyes}, R.~W., {Fischer}, D.~A., {Johnson}, J.~A.,
  {Marcy}, G.~W., {Howard}, A.~W., {Szentgyorgyi}, A., {F{\H u}r{\'e}sz}, G.,
  {Buchhave}, L.~A., {B{\'e}ky}, B., {Sasselov}, D.~D., {Stefanik}, R.~P.,
  {Perumpilly}, G., {Everett}, M., {L{\'a}z{\'a}r}, J., {Papp}, I., \&
  {S{\'a}ri}, P. 2012, \apj, 745, 80

\bibitem[{{Sanchis-Ojeda} {et~al.}(2011){Sanchis-Ojeda}, {Winn}, {Holman},
  {Carter}, {Osip}, \& {Fuentes}}]{Sanchis-Ojeda_et_al_11}
{Sanchis-Ojeda}, R., {Winn}, J.~N., {Holman}, M.~J., {Carter}, J.~A., {Osip},
  D.~J., \& {Fuentes}, C.~I. 2011, \apj, 733, 127

\bibitem[{{Savonije} \& {Papaloizou}(1997)}]{Savonije_Papaloizou_97b}
{Savonije}, G.~J. \& {Papaloizou}, J.~C.~B. 1997, \mnras, 291, 633

\bibitem[{{Scharlemann}(1981)}]{Scharlemann_81}
{Scharlemann}, E.~T. 1981, \apj, 246, 292

\bibitem[{{Scharlemann}(1982)}]{Scharlemann_82}
---. 1982, \apj, 253, 298

\bibitem[{{Schlaufman} {et~al.}(2010){Schlaufman}, {Lin}, \&
  {Ida}}]{Schlaufman_Lin_Ida_10}
{Schlaufman}, K.~C., {Lin}, D.~N.~C., \& {Ida}, S. 2010, \apjl, 724, L53

\bibitem[{{Simpson} {et~al.}(2011){Simpson}, {Faedi}, {Barros}, {Brown},
  {Collier Cameron}, {Hebb}, {Pollacco}, {Smalley}, {Todd}, {Butters},
  {H{\'e}brard}, {McCormac}, {Miller}, {Santerne}, {Street}, {Skillen},
  {Triaud}, {Anderson}, {Bento}, {Boisse}, {Bouchy}, {Enoch}, {Haswell},
  {Hellier}, {Holmes}, {Horne}, {Keenan}, {Lister}, {Maxted}, {Moulds},
  {Moutou}, {Norton}, {Parley}, {Pepe}, {Queloz}, {Segransan}, {Smith},
  {Stempels}, {Udry}, {Watson}, {West}, \& {Wheatley}}]{Simpson_et_al_11}
{Simpson}, E.~K., {Faedi}, F., {Barros}, S.~C.~C., {Brown}, D.~J.~A., {Collier
  Cameron}, A., {Hebb}, L., {Pollacco}, D., {Smalley}, B., {Todd}, I.,
  {Butters}, O.~W., {H{\'e}brard}, G., {McCormac}, J., {Miller}, G.~R.~M.,
  {Santerne}, A., {Street}, R.~A., {Skillen}, I., {Triaud}, A.~H.~M.~J.,
  {Anderson}, D.~R., {Bento}, J., {Boisse}, I., {Bouchy}, F., {Enoch}, B.,
  {Haswell}, C.~A., {Hellier}, C., {Holmes}, S., {Horne}, K., {Keenan}, F.~P.,
  {Lister}, T.~A., {Maxted}, P.~F.~L., {Moulds}, V., {Moutou}, C., {Norton},
  A.~J., {Parley}, N., {Pepe}, F., {Queloz}, D., {Segransan}, D., {Smith},
  A.~M.~S., {Stempels}, H.~C., {Udry}, S., {Watson}, C.~A., {West}, R.~G., \&
  {Wheatley}, P.~J. 2011, \aj, 141, 8

\bibitem[{{Skillen} {et~al.}(2009){Skillen}, {Pollacco}, {Collier Cameron},
  {Hebb}, {Simpson}, {Bouchy}, {Christian}, {Gibson}, {H{\'e}brard}, {Joshi},
  {Loeillet}, {Smalley}, {Stempels}, {Street}, {Udry}, {West}, {Anderson},
  {Barros}, {Enoch}, {Haswell}, {Hellier}, {Horne}, {Irwin}, {Keenan},
  {Lister}, {Maxted}, {Mayor}, {Moutou}, {Norton}, {Parley}, {Queloz}, {Ryans},
  {Todd}, {Wheatley}, \& {Wilson}}]{Skillen_et_al_09}
{Skillen}, I., {Pollacco}, D., {Collier Cameron}, A., {Hebb}, L., {Simpson},
  E., {Bouchy}, F., {Christian}, D.~J., {Gibson}, N.~P., {H{\'e}brard}, G.,
  {Joshi}, Y.~C., {Loeillet}, B., {Smalley}, B., {Stempels}, H.~C., {Street},
  R.~A., {Udry}, S., {West}, R.~G., {Anderson}, D.~R., {Barros}, S.~C.~C.,
  {Enoch}, B., {Haswell}, C.~A., {Hellier}, C., {Horne}, K., {Irwin}, J.,
  {Keenan}, F.~P., {Lister}, T.~A., {Maxted}, P., {Mayor}, M., {Moutou}, C.,
  {Norton}, A.~J., {Parley}, N., {Queloz}, D., {Ryans}, R., {Todd}, I.,
  {Wheatley}, P.~J., \& {Wilson}, D.~M. 2009, \aap, 502, 391

\bibitem[{{Smalley} {et~al.}(2010){Smalley}, {Anderson}, {Collier Cameron},
  {Gillon}, {Hellier}, {Lister}, {Maxted}, {Queloz}, {Triaud}, {West},
  {Bentley}, {Enoch}, {Pepe}, {Pollacco}, {Segransan}, {Smith}, {Southworth},
  {Udry}, {Wheatley}, {Wood}, \& {Bento}}]{Smalley_et_al_10}
{Smalley}, B., {Anderson}, D.~R., {Collier Cameron}, A., {Gillon}, M.,
  {Hellier}, C., {Lister}, T.~A., {Maxted}, P.~F.~L., {Queloz}, D., {Triaud},
  A.~H.~M.~J., {West}, R.~G., {Bentley}, S.~J., {Enoch}, B., {Pepe}, F.,
  {Pollacco}, D.~L., {Segransan}, D., {Smith}, A.~M.~S., {Southworth}, J.,
  {Udry}, S., {Wheatley}, P.~J., {Wood}, P.~L., \& {Bento}, J. 2010, \aap, 520,
  A56+

\bibitem[{{Smalley} {et~al.}(2011){Smalley}, {Anderson}, {Collier Cameron},
  {Hellier}, {Lendl}, {Maxted}, {Queloz}, {Triaud}, {West}, {Bentley}, {Enoch},
  {Gillon}, {Lister}, {Pepe}, {Pollacco}, {Segransan}, {Smith}, {Southworth},
  {Udry}, {Wheatley}, {Wood}, \& {Bento}}]{Smalley_et_al_11}
{Smalley}, B., {Anderson}, D.~R., {Collier Cameron}, A., {Hellier}, C.,
  {Lendl}, M., {Maxted}, P.~F.~L., {Queloz}, D., {Triaud}, A.~H.~M.~J., {West},
  R.~G., {Bentley}, S.~J., {Enoch}, B., {Gillon}, M., {Lister}, T.~A., {Pepe},
  F., {Pollacco}, D., {Segransan}, D., {Smith}, A.~M.~S., {Southworth}, J.,
  {Udry}, S., {Wheatley}, P.~J., {Wood}, P.~L., \& {Bento}, J. 2011, \aap, 526,
  A130+

\bibitem[{{Smith} {et~al.}(2006){Smith}, {Collier Cameron}, {Christian},
  {Clarkson}, {Enoch}, {Evans}, {Haswell}, {Hellier}, {Horne}, {Irwin}, {Kane},
  {Lister}, {Norton}, {Parley}, {Pollacco}, {Ryans}, {Skillen}, {Street},
  {Triaud}, {West}, {Wheatley}, \& {Wilson}}]{Smith_et_al_06}
{Smith}, A.~M.~S., {Collier Cameron}, A., {Christian}, D.~J., {Clarkson},
  W.~I., {Enoch}, B., {Evans}, A., {Haswell}, C.~A., {Hellier}, C., {Horne},
  K., {Irwin}, J., {Kane}, S.~R., {Lister}, T.~A., {Norton}, A.~J., {Parley},
  N., {Pollacco}, D.~L., {Ryans}, R., {Skillen}, I., {Street}, R.~A., {Triaud},
  A.~H.~M.~J., {West}, R.~G., {Wheatley}, P.~J., \& {Wilson}, D.~M. 2006,
  \mnras, 373, 1151

\bibitem[{{Snellen}(2004)}]{Snellen_04}
{Snellen}, I.~A.~G. 2004, \mnras, 353, L1

\bibitem[{{Southworth}(2010)}]{Southworth_10}
{Southworth}, J. 2010, \mnras, 408, 1689

\bibitem[{{Southworth} {et~al.}(2009{\natexlab{a}}){Southworth}, {Hinse},
  {Burgdorf}, {Dominik}, {Hornstrup}, {J{\o}rgensen}, {Liebig}, {Ricci},
  {Th{\"o}ne}, {Anguita}, {Bozza}, {Novati}, {Harps{\o}e}, {Mancini}, {Masi},
  {Mathiasen}, {Rahvar}, {Scarpetta}, {Snodgrass}, {Surdej}, \&
  {Zub}}]{Southworth_et_al_09a}
{Southworth}, J., {Hinse}, T.~C., {Burgdorf}, M.~J., {Dominik}, M.,
  {Hornstrup}, A., {J{\o}rgensen}, U.~G., {Liebig}, C., {Ricci}, D.,
  {Th{\"o}ne}, C.~C., {Anguita}, T., {Bozza}, V., {Novati}, S.~C.,
  {Harps{\o}e}, K., {Mancini}, L., {Masi}, G., {Mathiasen}, M., {Rahvar}, S.,
  {Scarpetta}, G., {Snodgrass}, C., {Surdej}, J., \& {Zub}, M.
  2009{\natexlab{a}}, \mnras, 399, 287

\bibitem[{{Southworth} {et~al.}(2009{\natexlab{b}}){Southworth}, {Hinse},
  {J{\o}rgensen}, {Dominik}, {Ricci}, {Burgdorf}, {Hornstrup}, {Wheatley},
  {Anguita}, {Bozza}, {Novati}, {Harps{\o}e}, {Kj{\ae}rgaard}, {Liebig},
  {Mancini}, {Masi}, {Mathiasen}, {Rahvar}, {Scarpetta}, {Snodgrass}, {Surdej},
  {Th{\"o}ne}, \& {Zub}}]{Southworth_et_al_09b}
{Southworth}, J., {Hinse}, T.~C., {J{\o}rgensen}, U.~G., {Dominik}, M.,
  {Ricci}, D., {Burgdorf}, M.~J., {Hornstrup}, A., {Wheatley}, P.~J.,
  {Anguita}, T., {Bozza}, V., {Novati}, S.~C., {Harps{\o}e}, K.,
  {Kj{\ae}rgaard}, P., {Liebig}, C., {Mancini}, L., {Masi}, G., {Mathiasen},
  M., {Rahvar}, S., {Scarpetta}, G., {Snodgrass}, C., {Surdej}, J.,
  {Th{\"o}ne}, C.~C., \& {Zub}, M. 2009{\natexlab{b}}, \mnras, 396, 1023

\bibitem[{{Sozzetti} {et~al.}(2007){Sozzetti}, {Torres}, {Charbonneau},
  {Latham}, {Holman}, {Winn}, {Laird}, \& {O'Donovan}}]{Sozzetti_et_al_07}
{Sozzetti}, A., {Torres}, G., {Charbonneau}, D., {Latham}, D.~W., {Holman},
  M.~J., {Winn}, J.~N., {Laird}, J.~B., \& {O'Donovan}, F.~T. 2007, \apj, 664,
  1190

\bibitem[{{Sozzetti} {et~al.}(2004){Sozzetti}, {Yong}, {Torres}, {Charbonneau},
  {Latham}, {Allende Prieto}, {Brown}, {Carney}, \& {Laird}}]{Sozzetti_04}
{Sozzetti}, A., {Yong}, D., {Torres}, G., {Charbonneau}, D., {Latham}, D.~W.,
  {Allende Prieto}, C., {Brown}, T.~M., {Carney}, B.~W., \& {Laird}, J.~B.
  2004, \apjl, 616, L167

\bibitem[{{Stauffer} \& {Hartmann}(1987)}]{Stauffer_Hartmann_87}
{Stauffer}, J.~R. \& {Hartmann}, L.~W. 1987, \apj, 318, 337

\bibitem[{{Street} {et~al.}(2010){Street}, {Simpson}, {Barros}, {Pollacco},
  {Joshi}, {Todd}, {Collier Cameron}, {Enoch}, {Parley}, {Stempels}, {Hebb},
  {Triaud}, {Queloz}, {Segransan}, {Pepe}, {Udry}, {Lister}, {Depagne}, {West},
  {Norton}, {Smalley}, {Hellier}, {Anderson}, {Maxted}, {Bentley}, {Skillen},
  {Gillon}, {Wheatley}, {Bento}, {Cathaway-Kjontvedt}, \&
  {Christian}}]{Street_et_al_10}
{Street}, R.~A., {Simpson}, E., {Barros}, S.~C.~C., {Pollacco}, D., {Joshi},
  Y., {Todd}, I., {Collier Cameron}, A., {Enoch}, B., {Parley}, N., {Stempels},
  E., {Hebb}, L., {Triaud}, A.~H.~M.~J., {Queloz}, D., {Segransan}, D., {Pepe},
  F., {Udry}, S., {Lister}, T.~A., {Depagne}, {\'E}., {West}, R.~G., {Norton},
  A.~J., {Smalley}, B., {Hellier}, C., {Anderson}, D.~R., {Maxted}, P.~F.~L.,
  {Bentley}, S.~J., {Skillen}, I., {Gillon}, M., {Wheatley}, P., {Bento}, J.,
  {Cathaway-Kjontvedt}, P., \& {Christian}, D.~J. 2010, \apj, 720, 337

\bibitem[{{Takeda} {et~al.}(2007){Takeda}, {Ford}, {Sills}, {Rasio}, {Fischer},
  \& {Valenti}}]{Takeda_et_al_07}
{Takeda}, G., {Ford}, E.~B., {Sills}, A., {Rasio}, F.~A., {Fischer}, D.~A., \&
  {Valenti}, J.~A. 2007, \apjs, 168, 297

\bibitem[{{Terquem} {et~al.}(1998){Terquem}, {Papaloizou}, {Nelson}, \&
  {Lin}}]{Terquem_et_al_98}
{Terquem}, C., {Papaloizou}, J.~C.~B., {Nelson}, R.~P., \& {Lin}, D.~N.~C.
  1998, \apj, 502, 788

\bibitem[{{Tripathi} {et~al.}(2010){Tripathi}, {Winn}, {Johnson}, {Howard},
  {Halverson}, {Marcy}, {Holman}, {de Kleer}, {Carter}, {Esquerdo}, {Everett},
  \& {Cabrera}}]{Tripathi_10}
{Tripathi}, A., {Winn}, J.~N., {Johnson}, J.~A., {Howard}, A.~W., {Halverson},
  S., {Marcy}, G.~W., {Holman}, M.~J., {de Kleer}, K.~R., {Carter}, J.~A.,
  {Esquerdo}, G.~A., {Everett}, M.~E., \& {Cabrera}, N.~E. 2010, \apj, 715, 421

\bibitem[{{West} {et~al.}(2010){West}, {Anderson}, {Brown}, {Collier Cameron},
  {Gillon}, {Hellier}, {Lister}, {Maxted}, {Queloz}, {Enoch}, {Parley}, {Pepe},
  {Pollacco}, {Segransan}, {Smalley}, {Triaud}, {Udry}, \&
  {Wheatley}}]{West_et_al_10}
{West}, R.~G., {Anderson}, D.~R., {Brown}, D.~J.~A., {Collier Cameron}, A.,
  {Gillon}, M., {Hellier}, C., {Lister}, T.~A., {Maxted}, P.~F.~L., {Queloz},
  D., {Enoch}, B., {Parley}, N.~R., {Pepe}, F., {Pollacco}, D., {Segransan},
  D., {Smalley}, B., {Triaud}, A.~H.~M.~J., {Udry}, S., \& {Wheatley}, P.~J.
  2010, ApJ, submitted

\bibitem[{{Wilson} {et~al.}(2008){Wilson}, {Gillon}, {Hellier}, {Maxted},
  {Pepe}, {Queloz}, {Anderson}, {Collier Cameron}, {Smalley}, {Lister},
  {Bentley}, {Blecha}, {Christian}, {Enoch}, {Haswell}, {Hebb}, {Horne},
  {Irwin}, {Joshi}, {Kane}, {Marmier}, {Mayor}, {Parley}, {Pollacco}, {Pont},
  {Ryans}, {Segransan}, {Skillen}, {Street}, {Udry}, {West}, \&
  {Wheatley}}]{Wilson_et_al_08}
{Wilson}, D.~M., {Gillon}, M., {Hellier}, C., {Maxted}, P.~F.~L., {Pepe}, F.,
  {Queloz}, D., {Anderson}, D.~R., {Collier Cameron}, A., {Smalley}, B.,
  {Lister}, T.~A., {Bentley}, S.~J., {Blecha}, A., {Christian}, D.~J., {Enoch},
  B., {Haswell}, C.~A., {Hebb}, L., {Horne}, K., {Irwin}, J., {Joshi}, Y.~C.,
  {Kane}, S.~R., {Marmier}, M., {Mayor}, M., {Parley}, N., {Pollacco}, D.,
  {Pont}, F., {Ryans}, R., {Segransan}, D., {Skillen}, I., {Street}, R.~A.,
  {Udry}, S., {West}, R.~G., \& {Wheatley}, P.~J. 2008, \apjl, 675, L113

\bibitem[{{Winn} {et~al.}(2011){Winn}, {Howard}, {Johnson}, {Marcy},
  {Isaacson}, {Shporer}, {Bakos}, {Hartman}, {Holman}, {Albrecht}, {Crepp}, \&
  {Morton}}]{Winn_et_al_11}
{Winn}, J.~N., {Howard}, A.~W., {Johnson}, J.~A., {Marcy}, G.~W., {Isaacson},
  H., {Shporer}, A., {Bakos}, G.~{\'A}., {Hartman}, J.~D., {Holman}, M.~J.,
  {Albrecht}, S., {Crepp}, J.~R., \& {Morton}, T.~D. 2011, \aj, 141, 63

\bibitem[{{Winn} {et~al.}(2009){Winn}, {Johnson}, {Albrecht}, {Howard},
  {Marcy}, {Crossfield}, \& {Holman}}]{Winn_et_al_09}
{Winn}, J.~N., {Johnson}, J.~A., {Albrecht}, S., {Howard}, A.~W., {Marcy},
  G.~W., {Crossfield}, I.~J., \& {Holman}, M.~J. 2009, \apjl, 703, L99

\bibitem[{{Winn} {et~al.}(2010{\natexlab{a}}){Winn}, {Johnson}, {Howard},
  {Marcy}, {Bakos}, {Hartman}, {Torres}, {Albrecht}, \&
  {Narita}}]{Winn_et_al_10a}
{Winn}, J.~N., {Johnson}, J.~A., {Howard}, A.~W., {Marcy}, G.~W., {Bakos},
  G.~{\'A}., {Hartman}, J., {Torres}, G., {Albrecht}, S., \& {Narita}, N.
  2010{\natexlab{a}}, \apj, 718, 575

\bibitem[{{Winn} {et~al.}(2010{\natexlab{b}}){Winn}, {Johnson}, {Howard},
  {Marcy}, {Isaacson}, {Shporer}, {Bakos}, {Hartman}, \&
  {Albrecht}}]{Winn_et_al_10b}
{Winn}, J.~N., {Johnson}, J.~A., {Howard}, A.~W., {Marcy}, G.~W., {Isaacson},
  H., {Shporer}, A., {Bakos}, G.~{\'A}., {Hartman}, J.~D., \& {Albrecht}, S.
  2010{\natexlab{b}}, \apjl, 723, L223

\bibitem[{{Winn} {et~al.}(2006){Winn}, {Johnson}, {Marcy}, {Butler}, {Vogt},
  {Henry}, {Roussanova}, {Holman}, {Enya}, {Narita}, {Suto}, \&
  {Turner}}]{Winn_et_al_06}
{Winn}, J.~N., {Johnson}, J.~A., {Marcy}, G.~W., {Butler}, R.~P., {Vogt},
  S.~S., {Henry}, G.~W., {Roussanova}, A., {Holman}, M.~J., {Enya}, K.,
  {Narita}, N., {Suto}, Y., \& {Turner}, E.~L. 2006, \apjl, 653, L69

\bibitem[{{Winn} {et~al.}(2008){Winn}, {Johnson}, {Narita}, {Suto}, {Turner},
  {Fischer}, {Butler}, {Vogt}, {O'Donovan}, \& {Gaudi}}]{Winn_et_al_08}
{Winn}, J.~N., {Johnson}, J.~A., {Narita}, N., {Suto}, Y., {Turner}, E.~L.,
  {Fischer}, D.~A., {Butler}, R.~P., {Vogt}, S.~S., {O'Donovan}, F.~T., \&
  {Gaudi}, B.~S. 2008, \apj, 682, 1283

\bibitem[{{Wolff} \& {Simon}(1997)}]{Wolff_Simon_97}
{Wolff}, S. \& {Simon}, T. 1997, \pasp, 109, 759

\bibitem[{{Wu}(2005{\natexlab{a}})}]{Wu_05a}
{Wu}, Y. 2005{\natexlab{a}}, \apj, 635, 674

\bibitem[{{Wu}(2005{\natexlab{b}})}]{Wu_05b}
---. 2005{\natexlab{b}}, \apj, 635, 688

\bibitem[{{Zahn}(1966)}]{Zahn_66}
{Zahn}, J.~P. 1966, Ann. d'Astrophys., 29, 489

\bibitem[{{Zahn}(1970)}]{Zahn_70}
---. 1970, \aap, 4, 452

\bibitem[{{Zahn}(1975)}]{Zahn_75}
{Zahn}, J.-P. 1975, \aap, 41, 329

\bibitem[{Zahn(1977)}]{Zahn_77}
Zahn, J.~P. 1977, \aap, 57, 383

\bibitem[{{Zahn} \& {Bouchet}(1989)}]{Zahn_89}
{Zahn}, J.-P. \& {Bouchet}, L. 1989, \aap, 223, 112

\end{thebibliography}
\bibliographystyle{apj}

\end{document}